\documentclass{aa}  

\usepackage{graphicx}
\usepackage{txfonts}
\newcommand{\degrees}{\ensuremath{^\circ}}

\begin{document} 

\titlerunning{Galaxies with KDCs in Illustris}
   \title{Galaxies with kinematically distinct cores in Illustris}

   \author{Ivana Ebrov\'{a} \inst{1}\thanks{\email{ebrova.ivana@gmail.com}}
          \and
          Ewa L. {\L}okas \inst{1}
          \and
          Ji\v{r}\'{i} Eli\'{a}\v{s}ek \inst{2,3}
          }

  \institute{Nicolaus Copernicus Astronomical Center, Polish Academy of Sciences, Bartycka 18, 00-716 Warsaw, Poland
        \and
         Institute of Theoretical Physics, Faculty of Mathematics and Physics, Charles University, V~Hole\v{s}ovi\v{c}k\'ach 2, 180\,00 Prague, Czech Republic
        \and
	FZU -- Institute of Physics of the Czech Academy of Sciences, Na Slovance 1999/2, Prague 182\,21, Czech Republic
         }

    \date{Received ; accepted }

 
  \abstract
   {
The growing amount of integral-field spectroscopic data creates an increased demand for understanding kinematic peculiarities that carry valuable information about the evolution of the host galaxies.
}
   {
For kinematically distinct cores (KDCs), a number of formation mechanisms have been proposed, but it is still unclear which of them commonly occur in the Universe.
We aim to address the KDC formation in the cosmological context.
}
   {
We used the publicly available data of the large-scale hydrodynamic cosmological simulation Illustris.
We identify 134 KDCs, study their properties, and follow their evolution back in time. 
Examples of four galaxies hosting KDCs are presented and described in detail. 
}
   {
The masses of the KDC hosts follow the general distribution of the Illustris galaxies, with a possible slight preference towards massive galaxies.
KDCs can be long-lived features, with their formation epochs roughly uniformly distributed in look-back times 0\,--\,11.4\,Gyr, and they can survive even major or multiple subsequent mergers.
There is no single channel of KDC formation, but mergers seem to be the formation mechanism for about 60\,\% of KDCs with a significant preference for major mergers and with the percentage being higher among massive hosts.
Other KDCs formed during a pericentric passage or flyby of another galaxy, by precession of a previously formed rapidly rotating core, or without an obvious external cause. 
The mean mass-weighted stellar age inside the KDC radius is either about the same as the look-back time of the KDC formation or older.
Although the radii of our KDCs are on average larger than observed, we find that younger stellar ages are typically associated with smaller KDCs. 
A significant fraction of KDC hosts possess stellar shells formed during mergers that led to KDCs within the last 5\,Gyr, or double peaks in their velocity dispersion maps.
}
   {}


   \maketitle
%

\section{Introduction} \label{sec:intro}

Peculiarities in galaxies carry valuable information on the evolution of the host galaxies. 
Different features can tell different pieces of the story about how their hosts were born and shaped over the course of time. 
In this paper, we focus on kinematically distinct cores (KDCs), i.e. central stellar components that show distinct properties of mean line-of-sight velocity from the main body of the host galaxy.

KDCs have been known for almost four decades, originally having been studied in one-dimensional (long-)slit spectroscopic data \citep{ef82,be88,fi88,js88}.
With the increasing availability of two-dimensional observations -- integral field spectroscopy (IFS), e.g. Spectrographic Areal Unit for Research on Optical Nebulae \citep[SAURON;][]{sauron}, ATLAS$^{3{\rm D}}$ \citep{a3d1}, Sydney-AAO Multiobject IFU \citep[SAMI;][]{sami}, Calar Alto Legacy Integral Field Area Survey \citep[CALIFA;][]{califa1}, and Mapping Nearby Galaxies at APO \citep[MaNGA;][]{manga} -- KDCs are being detected in larger quantities.

As is the case with many other objects and features in galactic astronomy, the term “KDC” is not well codified and the relevant features are subject to different terminology and classifications throughout the literature.
They can be called \textit{kinematically decoupled cores}, \textit{kinematically peculiar cores}, or just \textit{kinematically distinct} (or \textit{decoupled}) \textit{components}.
They are divided and included into different categories. 
The IFS data are often analyzed with the help of the \textsc{kinemetry} routine \citep{kinemetry}, most notably the SAURON and ATLAS$^{3{\rm D}}$ samples.

Four KDC hosts of the SAURON survey, reported in \cite{mcd06}, are not labeled as KDCs in the subsequent papers \citep{saur9,saur12}, instead they form their own category of \textit{central low-level velocity} (CLV) systems.
\cite{saur9} and \cite{saur12} defined \textit{kinematically decoupled component} as a stellar velocity component having either an abrupt change in the kinematic axis with a difference
larger than $20\degrees$, or an outer low-level velocity component (i.e. maximum velocity amplitude lower than 15\,km\,s$^{-1}$).

\cite{a3d2}, analyzing the ATLAS$^{3{\rm D}}$ sample, classified \textit{kinematically distinct core} as having an abrupt change of the kinematic axis by at least $30\degrees$, however, if the change is close to $180\degrees$, the feature is placed into a separate category of \textit{counter-rotating cores} (CRC). 
Moreover, in the category \textit{double $\sigma$} ($2\sigma$) -- two symmetric velocity dispersion peaks along the major axis (see also Sect.\,\ref{sec:dis}) -- most of the galaxies also have KDCs.

\cite{manga4}, inspecting the MaNGA sample, recognized two categories: \textit{kinematically decoupled cores} -- rotating cores that are small compared to the galaxy, which itself is non-rotating and $2\sigma$ \textit{counter-rotators} that show either two strong velocity dispersion peaks along the major axis, or clear counter-rotation and high velocity dispersion at the counter-rotation boundary. 

Somewhat in contrast to the last, \cite{sch18}, classifying galaxies in the Magneticum Pathfinder simulation, define \textit{distinct cores} as central rotating components surrounded by a low-level or non-rotating component and \textit{kinematically distinct cores} (including counter-rotating cores) as central rotating components surrounded by a region with inclined rotation with respect to the central component.
Throughout this paper, we include all the above listed groups under the abbreviation KDC, except for the $2\sigma$ category, where we consider only galaxies with the KDC visible in the mean line-of-sight velocity data.

KDCs are typically found in early-type galaxies (ETGs, i.e.  ellipticals and lenticulars). 
Based on their specific stellar angular momentum, ETGs are divided into two classes -- fast and slow rotators \citep[][see also \citealt{capp16} for a review]{saur9,saur12,a3d3}.
Fast rotators (FRs) are more numerous \citep[86\,\% of the ATLAS$^{3{\rm D}}$ sample;][]{a3d3} and have regular velocity fields which qualitatively resemble those of disks. 
Slow rotators (SRs) are more pressure supported and tend to be more massive and round. 
SRs are also more likely to host a KDC.
KDCs are found in 29\,\% of the galaxies (17\,\% of FRs and 67\,\% of SRs) in the SAURON survey of 48 nearby ETGs \citep{saur9,saur12}.
Counting all KDCs listed in \cite{mcd06}, \cite{saur9}, and \cite{a3d2}, the KDC hosts constitute only around 13\,\% of galaxies (around 5\,\% of FRs and over 60\,\% of SRs) in the ATLAS$^{3{\rm D}}$ project -- a complete volume- and magnitude-limited sample of 260 ETGs. 
\cite{ricci16} found 1 KDC in their IFS data of 10 massive nearby ETGs (10\,\%) and \cite{kz00} reported 32\,\% KDC rate in the long-slit spectra of 53 ETGs.

The incidence of KDCs does not show an obvious dependence on the density of the environment. 
Long-slit spectra of 13 isolated early-type galaxies reveal 4 (31\,\%) KDCs \citep{hau06}.
In the Fornax3D project -- an IFS survey of 31 brightest galaxies inside the virial radius of the Fornax cluster -- \cite{f3d} found 33\,\% of their 21 ETGs (32\,\% of FRs and 50\,\% of SRs) to host KDCs. 
In addition to this, they found one KDC among their 10 late-type galaxies (LTGs, i.e. spirals and irregulars) and thus the overall abundance of KDCs in their sample is 26\,\%.
\cite{LouPhd} reported 15 out of 49 brightest cluster galaxies (31\,\%) hosting KDCs.
MUSE Most Massive Galaxies \citep[M3G;][]{m3g} IFS survey uncovers 20\,\% of KDCs among 25 massive ETGs located in the densest galaxy environments.

However, outside the densest galaxy environments, KDCs seem to be underrepresented among the very massive ETGs.
There are only 2\,\% of galaxies hosting KDCs in the MASSIVE survey -- a volume-limited IFS study of 90 most massive ETGs \citep{massive10}. 
This is in contradiction with the previous findings that KDCs are more frequent among SRs and SRs are mostly massive galaxies. 
Indeed, SRs (including galaxies classified as non-rotators) constitute 76\,\% of the MASSIVE survey, yet there are only two KDCs reported in the whole sample.

KDCs are also found at a lower rate in early-type dwarf galaxies (dEs) without an evident dependence on the density of the environment. 
\cite{janz17} found one out of nine isolated dEs hosting a KDC, which is comparable to two KDCs reported by \cite{V1dE} in the sample of 39 dEs in the Virgo cluster. 
In groups or lower density environments, KDCs are detected in 2 out of 14 dEs \citep[see][and references therein]{V1dE}. 
KDCs were found in several spirals, i.e. late-type disk galaxies, in the form of a structure that counter-rotates with respect to the main stellar body and often has a disc-like morphology  \citep[e.g.][]{mk94,ver07,cocc11,f3d}. 
In spirals, KDCs occur significantly less frequently compared to lenticulars, i.e. early-type disk galaxies, where they were detected in less than 10\,\% of cases \citep{kfm96,pizz04,kann01}.

The KDCs of disk galaxies are typically characterized by younger stellar populations accompanied by a substantial amount of gas \citep{joh13,pizz14,mitz17}, while SRs predominantly host KDCs that show no stellar population signatures over and above the well-known metallicity gradients in early-type galaxies and are largely consistent with old (10\,Gyr or older) stellar populations \citep{saur17}.
\cite{mcd06} recognized two populations of KDCs in the IFS data of 14 ETGs hosting KDCs: 
(1) kpc-scale KDCs with diameters around 1\,kpc or larger and stellar populations older than 8\,Gyr, residing in SRs; 
(2) compact KDCs with diameters of less than a few hundred parsecs, showing a whole range of stellar ages, found exclusively in FRs, with kinematics close to counter-rotation around the same axis as their host.

\begin{figure*} [!htb]
\resizebox{\hsize}{!}{\includegraphics{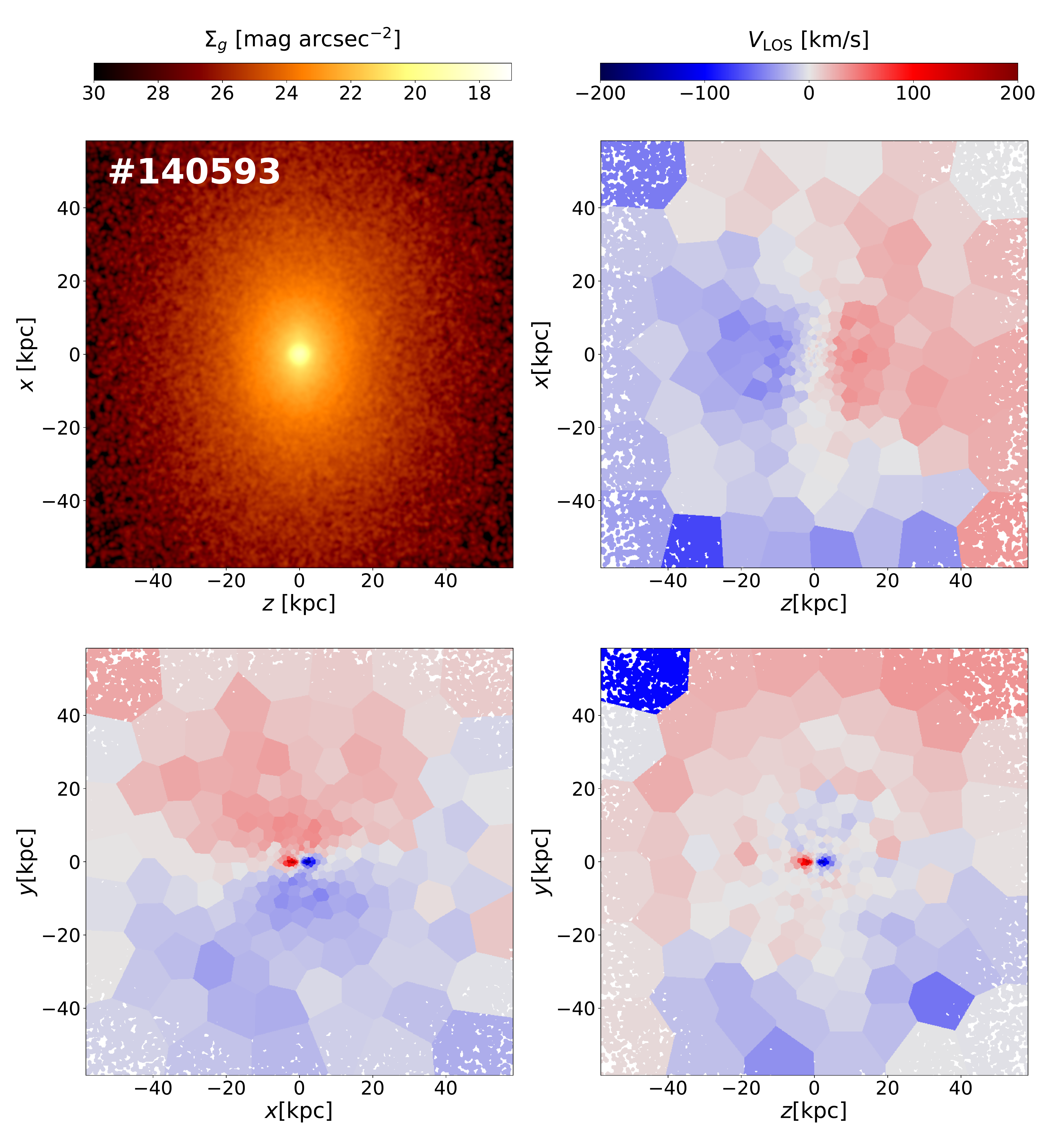}\hspace{3 cm}\includegraphics{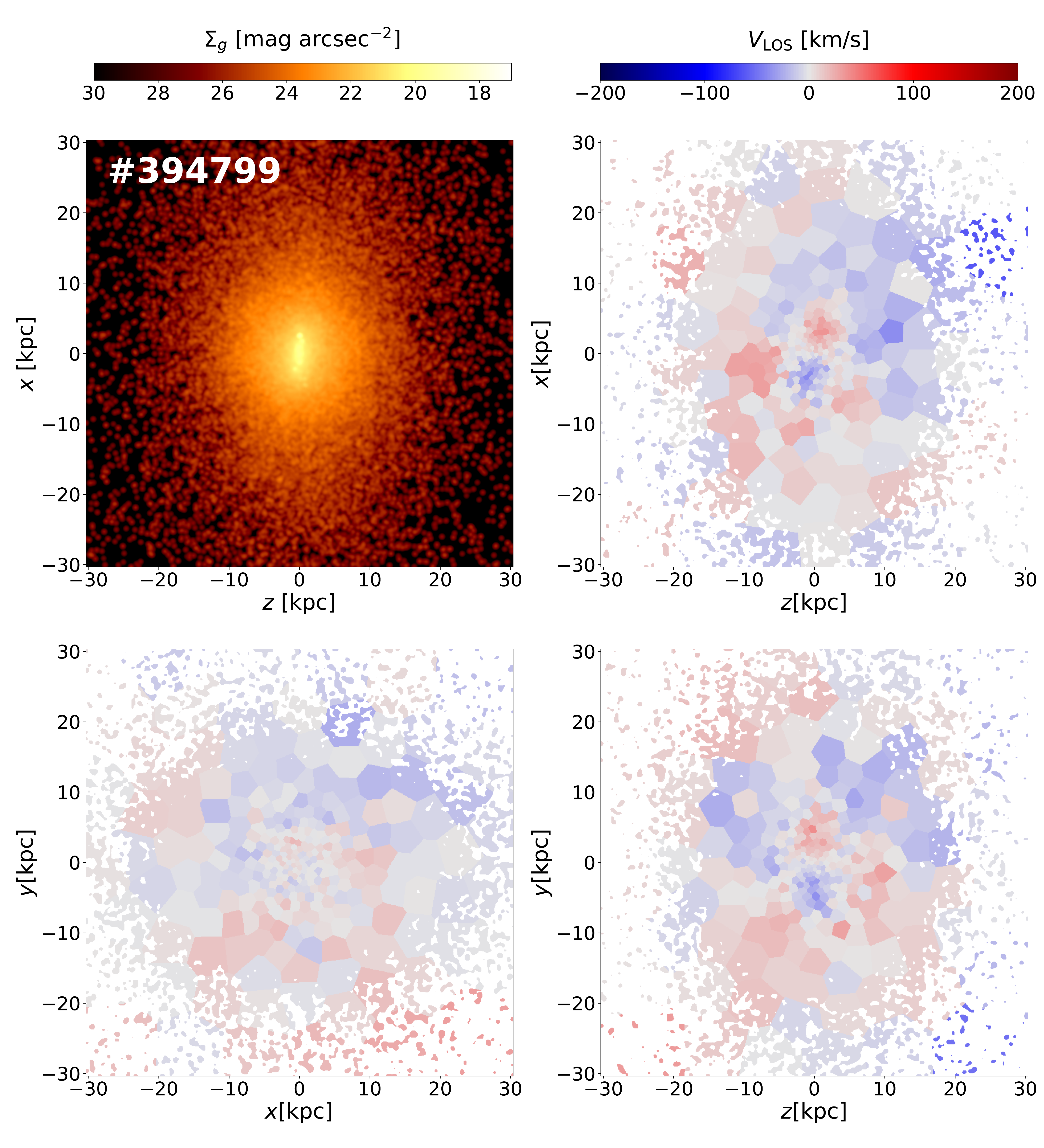}}
\resizebox{\hsize}{!}{\includegraphics{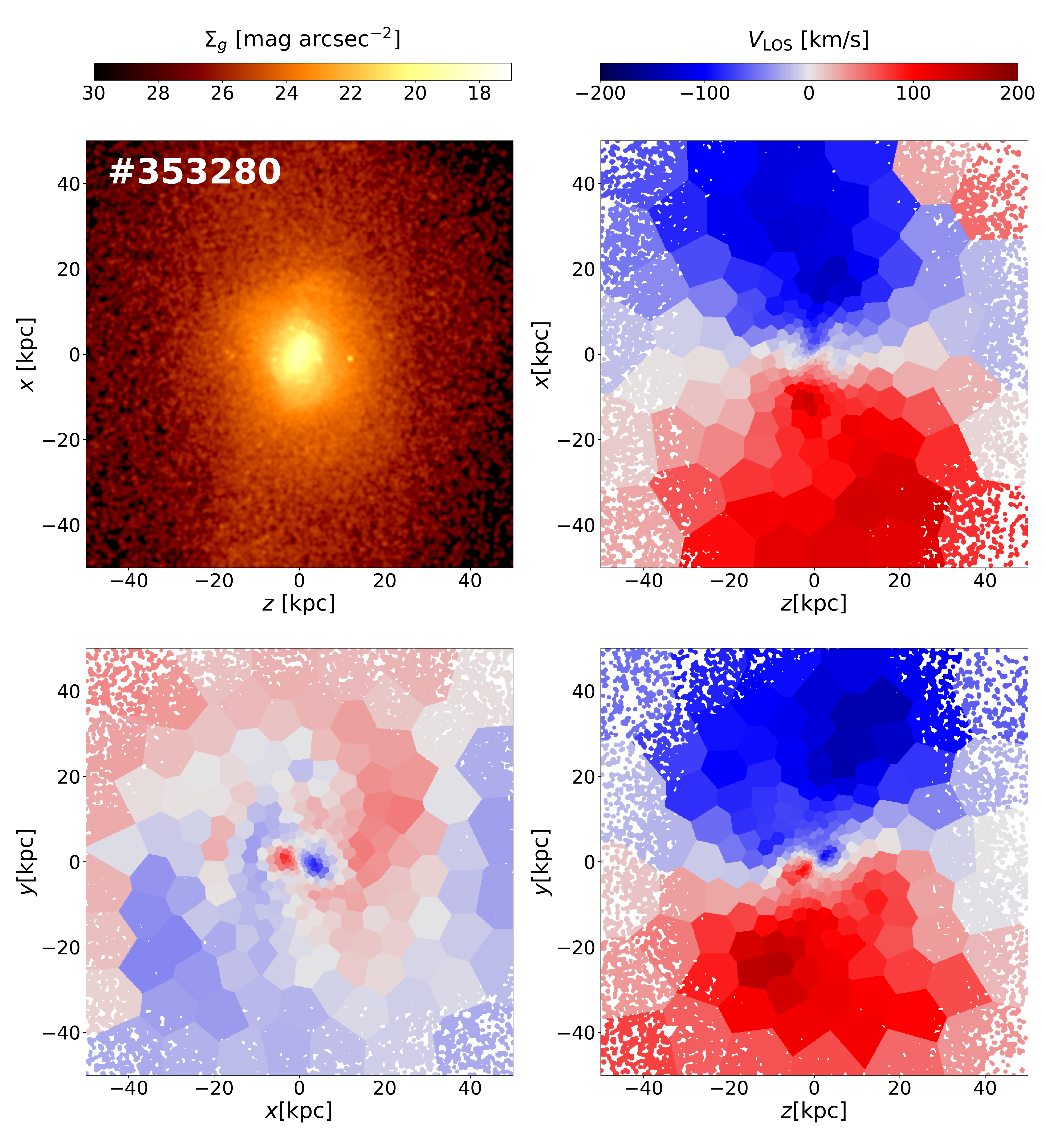}\hspace{3 cm}\includegraphics{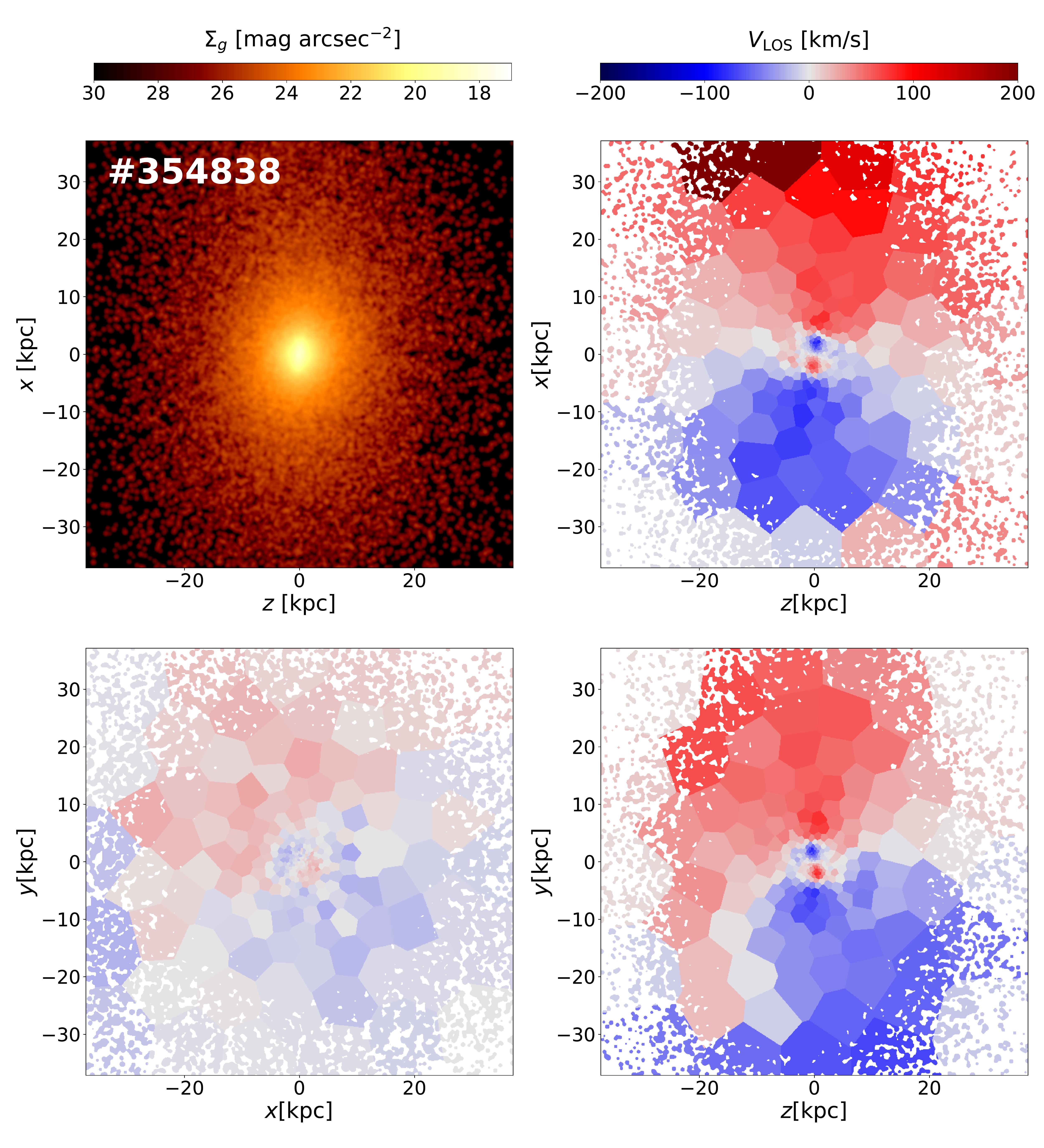}}
\caption{
Examples of four Illustris galaxies with KDCs from our sample (properties of the galaxies are listed in Tab.\,\ref{tab:ex}). 
Each $2\times2$ segment of panels belongs to one galaxy with the maps of surface density and mean line-of-sight velocity viewed along the intermediate axis of the galaxy displayed in the top left and right panels, respectively. 
Lower left and right panels show the mean line-of-sight velocity viewed along the minor and major axis, respectively.
The fields of view correspond to $1.5\times1.5\,r_{\rm max}$.
\label{fig:ex}}
\end{figure*}

Attempts to explain the origin of the KDCs are almost as old as their first observations.  
Already \cite{kor84} proposed that more robust cores of dEs can survive a merger with a bigger elliptical galaxy and form a KDC in its center. 
\cite{bq90} verified this scenario using self-consistent dissipationless (i.e. gas free) numerical simulations where the KDC is formed in retrograde mergers between elliptical galaxies of unequal mass (mass ratios 1:5 and 1:10).

After that, the focus shifted towards disk progenitors and more major mergers. 
KDCs were reproduced in dissipationless merger simulations of two disk progenitors -- \cite{bg98} reached the best results with mass ratio 1:2, while \cite{bb00} favors 1:1 mergers --  but major mergers with non-zero gas content seem to be more successful. 
\cite{jess07} explored 1:1 and 1:3 disk mergers with 0 and 0.1 gas fraction and found KDCs almost exclusively formed in equal-mass mergers with a dissipational component. 
\cite{hof10} investigated 1:1 disk mergers with gas fractions 0\,--\,0.4 and found KDCs only in remnants with at least 0.15 initial gas fraction and most of them in the range 0.15\,--\,0.2.
\cite{a3d6} simulated disk mergers with 0.1 gas fraction and remergers of the ETG merger products and concluded that the major mergers of disks (1:1 and 2:1) on retrograde orbits are most likely to produce KDCs. 
All their KDCs were destroyed when going through the subsequent remerger. 
\cite{sch17} studied a KDC produced in a retrograde 1:1 spiral galaxy merger simulation with 0.2 gas fraction. The rotation axis of the KDC engaged in a precession motion and the KDC dissolved after about 3\,Gyr. 

Generally, a retrograde configuration of the merger seems favorable to the KDC formation \citep{bq90,a3d6,sch17}, but \cite{tsa15} showed that KDCs can also originate from an initially prograde dissipational 1:1 merger, where the orbital spin is reversed by reactive forces due to substantial mass loss.
\cite{ran19} simulated simultaneously the origin of KDCs and cored density profiles of the ETGs centers. 
These observed cored profiles are difficult to obtain in models where the KDC formation relies on gas dissipation in galaxy centers resulting in star formation. 
\cite{ran19} performed a series of minor and major mergers with supermassive black holes (SMBHs) and the cored profile is a result of the scouring of stars in the centers by merging SMBHs. 
KDCs arise naturally from the same process in the major-merger simulations.
As opposed to the popular single major-merger origin, in a series of 95 hydrodynamical simulations of minor and major binary mergers, binary merger trees with multiple progenitors, and multiple sequential mergers of \cite{moo14}, KDCs are found infrequently in binary mergers, but multiple merger remnants commonly host KDCs.

KDCs were also studied in the cosmological context. 
\cite{a3d8} took advantage of semi-analytic modeling and concluded that KDCs formed in SRs during gas-rich major mergers at high redshift, followed by minor mergers, which build-up the outer layers of the remnants, while FRs are less likely to form KDCs since they underwent, on average, less than one major merger in their past.
In a set of cosmological hydrodynamical simulations \textit{Magneticum Pathfinder}, \cite{sch18} detected KDCs in 40 out of 900 galaxies (5\,\%), in similar rates among fast and slow rotators (4\,\% of FRs and 6\,\% of SRs). 

In another hydrodynamical cosmological simulation, \cite{tay18} found 2 out of 82 well-resolved galaxies having KDCs (2\,\%). None of the KDCs and other three kinematically atypical galaxies in their simulation formed due to major mergers. 
Originally field galaxies entered cosmological filaments where an accretion of gas with different angular momentum triggered AGN feedback.
The KDCs are formed in subsequent minor mergers on retrograde orbits, compared to the spin of the stellar component of the primary.
Similar mechanism of KDC formation -- consequential gas accretion from two distinct large-scale cosmological filaments in a specific spatial configuration -- was reported in a zoom-in cosmological simulation by \cite{al14}. 

Other merger-free scenarios of KDC origins include an early collapse without subsequent mergers \citep{hv98}, flybys of other galaxies \citep{ht94,ga05}, or a separatrix crossing mechanism which can, in case of a change of the symmetry of the gravitational potential, cause stars on box orbits to separate into clockwise and counterclockwise streaming tube orbits \citep{ec94}.

KDCs seen in the velocity maps can represent intrinsically more centrally concentrated embedded small components or just a part of more extended kinematic structures. 
Two kinematically different, but spatially overlapping components can leave their imprint in the form of enhanced velocity dispersion like the $2\sigma$ structure or $\sigma$-drop (see also Fig.\,\ref{fig:dispex}).
However, they can manifest themselves as a KDC in the mean line-of-sight velocity field only if one component is dominating the field in the center and the other in outer parts. 

Moreover, some KDCs can be not truly distinct/decoupled structures. 
The (weak) triaxial nature of the slow rotators \citep{saur10} allows them to support multiple types of different orbit families and, in some projections, these can appear as if the host possessed a KDC \citep{sta91}.
\cite{vdb08} demonstrated this effect with triaxial Schwarzschild modeling \citep{schw79} on a known KDC host NGC\,4365. 
They found no major transition in the orbital structure at the boundary of the KDC, meaning that the observed KDC is not physically distinct from the main body, but rather caused by a projected superposition of the orbits.
\cite{ned19} supported that idea with a kinematic decomposition technique applied on the IFS observation of the galaxy. 

Another effect that can cause a KDC to appear in the velocity maps is described in \cite{dd13,dd16}. 
Warps that commonly occur in the disks of galaxies, when observed face-on, can make an impression of a counter-rotating core in the projected velocity maps, even though the rotation axes of inner and outer parts of the disk are only slightly misaligned. 

In this work, we investigate the origin and properties of KDCs simulated in the cosmological context within the Illustris project \citep{vog14illpreintro,nel15illpub} -- a state-of-the-art large-scale ($106.5^{3}$\,Mpc$^{3}$ periodic box) cosmological simulation that follows the evolution of dark matter and different baryonic components from redshift $z=127$ up to the present time, using AREPO moving-mesh code \citep{sp10}. 
We use only the Illustris-1 run, which has $6\times10^{9}$ initial hydrodynamic cells with the mean mass of a baryonic particle $1.6\times10^{6}$\,M$_{\sun}$ at the end of the simulation and $6\times10^{9}$ dark matter particles with a particle mass $6.3\times10^{6}$\,M$_{\sun}$.
The particle data for 134 snapshots between redshifts 14 and 0 are publicly available, as well as subhalo catalogs and merger trees \citep{rg15illmer}.

The paper is organized as follows. 
In Sect.\,\ref{sec:selection} we describe the sample and our selection procedure, 
in Sects.~\ref{sec:birth} and \ref{sec:merger} we investigate the birth time of the KDCs and the merger history of their hosts, 
in Sects.~\ref{sec:age} and \ref{sec:corr} relations of measured quantities are examined, 
in Sect.\,\ref{sec:ex} four examples of the Illustris KDCs are shown in detail, and the results are discussed and summarized in Sects.~\ref{sec:dis} and \ref{sec:con}, respectively.

\begin{figure*} [!htb]
\resizebox{\hsize}{!}{\includegraphics{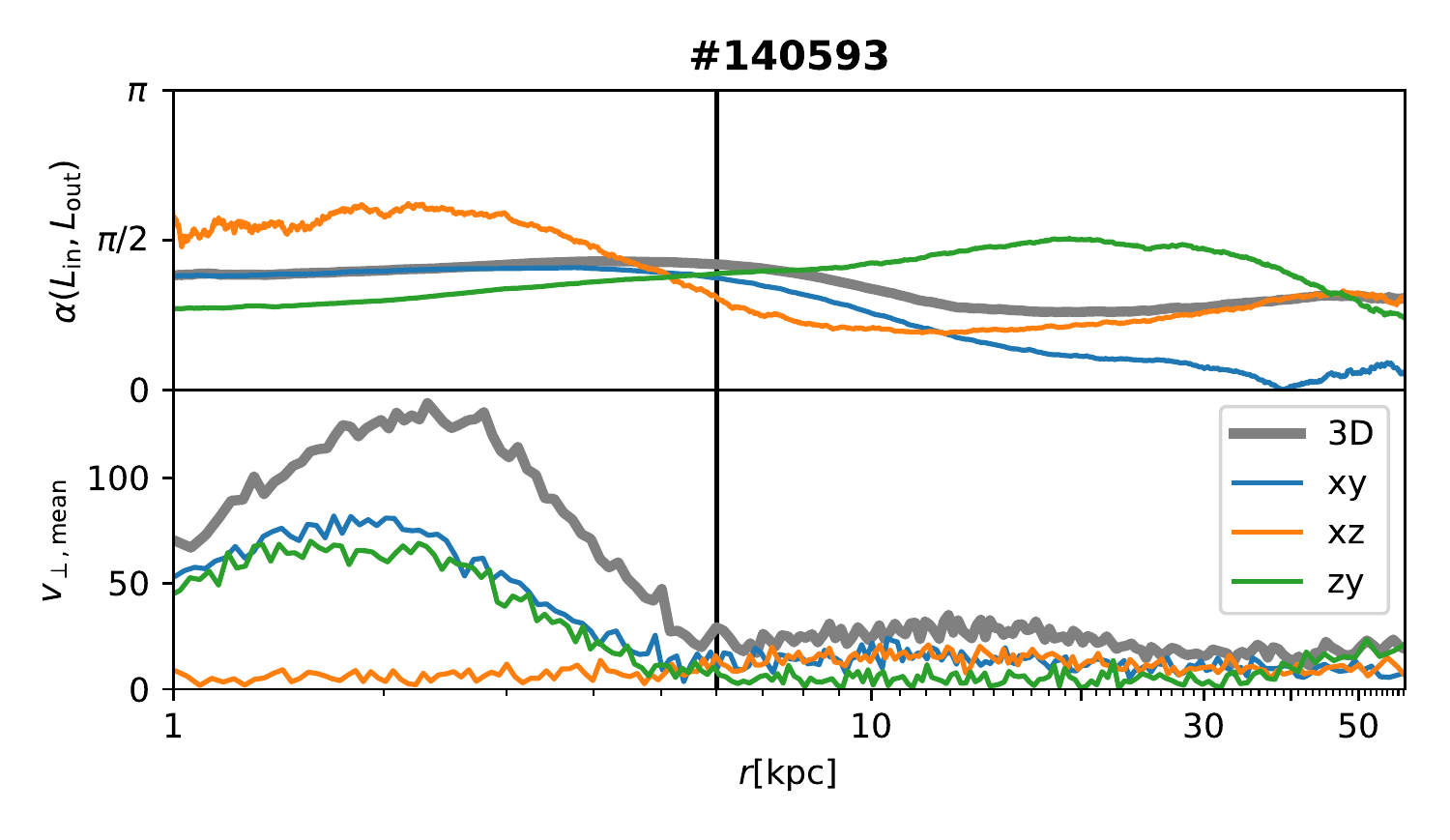}\includegraphics{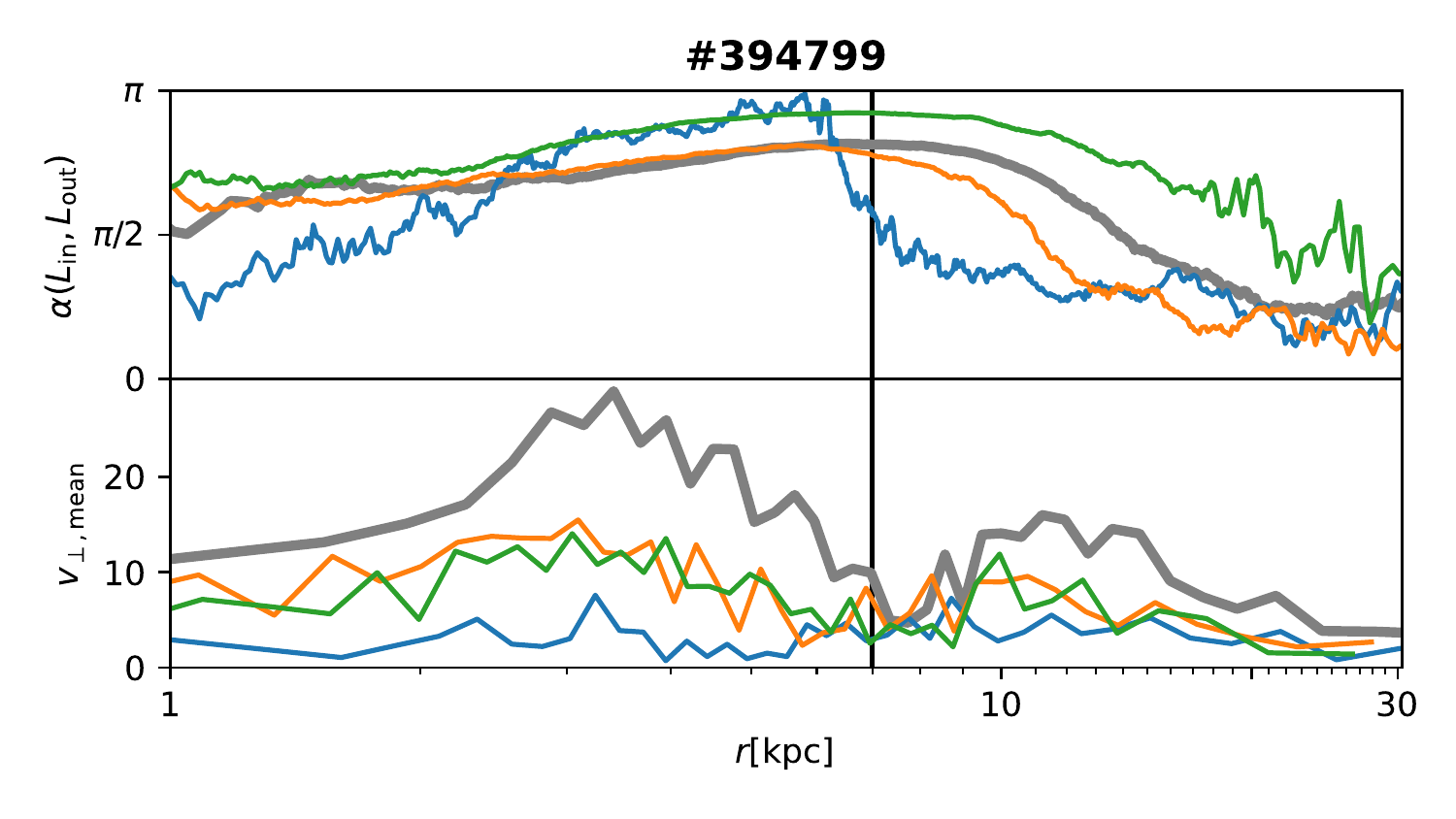}}
\resizebox{\hsize}{!}{\includegraphics{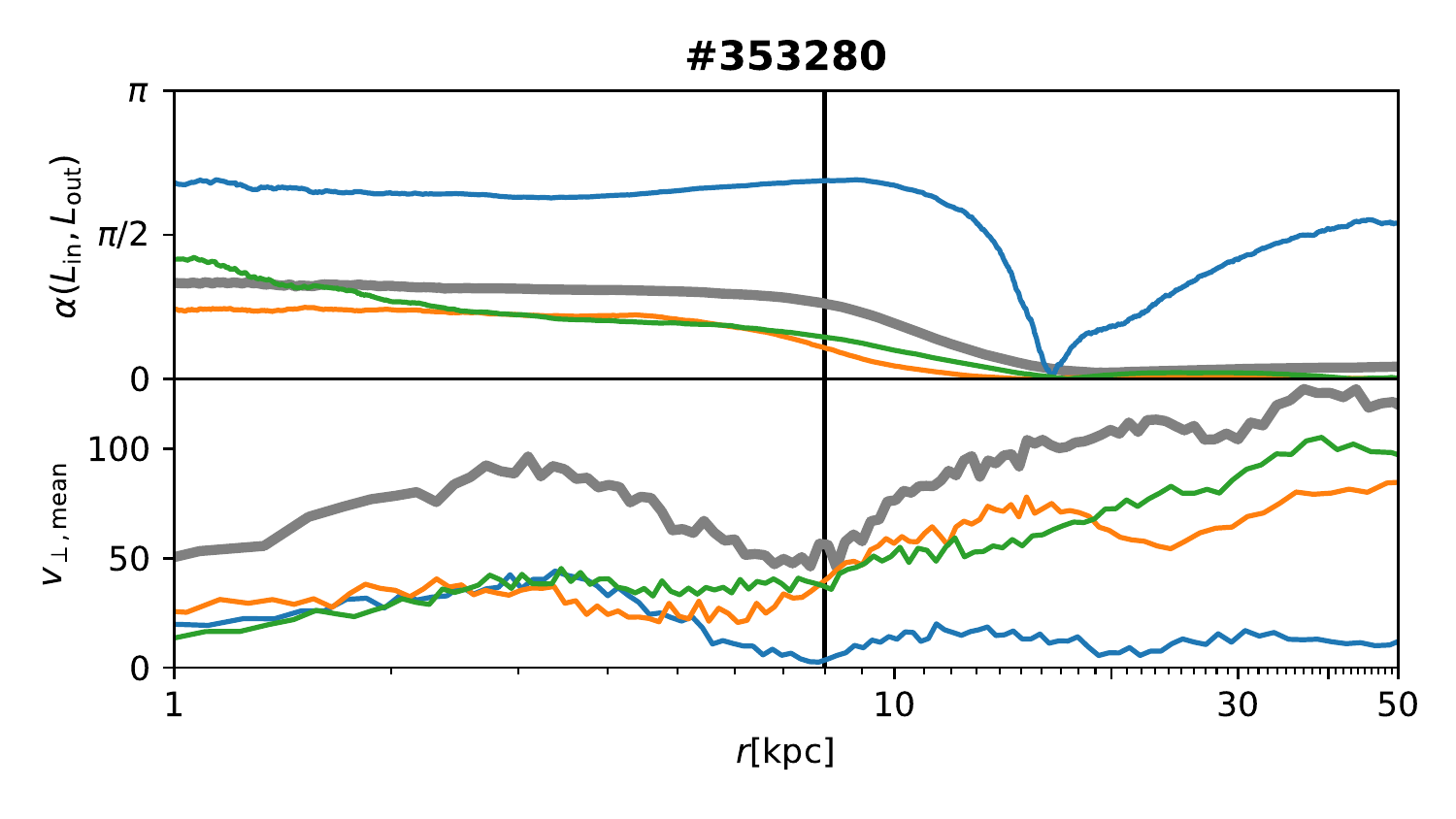}\includegraphics{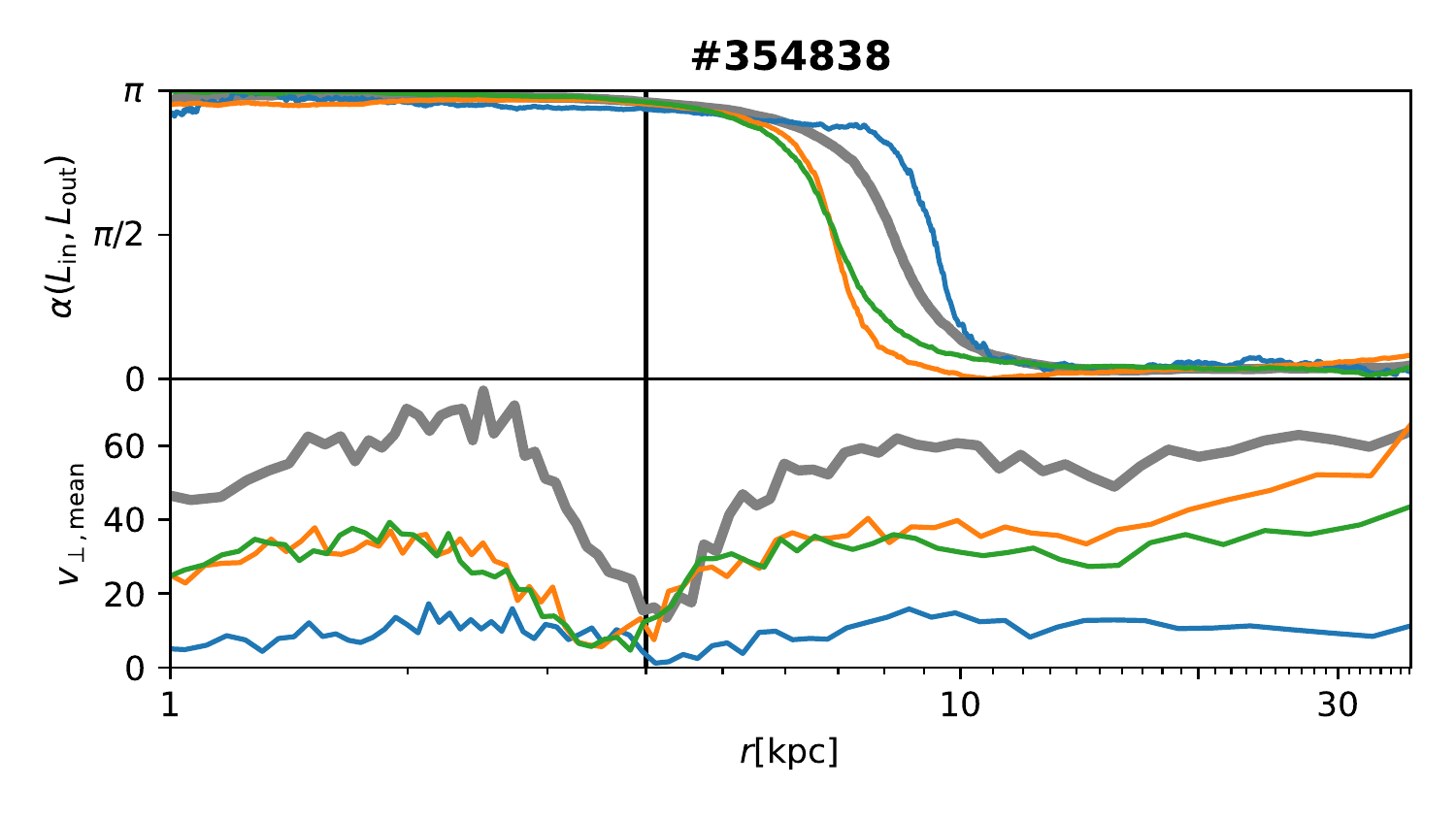}}
\caption{
Radial profiles of the mean tangential velocity, $v_{\perp,\rm mean}$, and the difference between the directions of the angular momenta inside and outside the given radius, $\alpha(L_{\rm in},L_{\rm out})$, for the stellar component of the four galaxies from Fig.\,\ref{fig:ex}. 
The quantities are measured in 3D and in the projection on three principal planes of the respective galaxy.
Angles are expressed in radians.
The black vertical lines denote the inferred KDC radius, $r_{\rm KDC}$.
\label{fig:cos12}}
\end{figure*}

\section{Results} \label{sec:res} 

\subsection{Sample selection and characteristics} \label{sec:selection}

Our sample of galaxies with kinematically distinct cores in Illustris was selected by visual inspection of maps of the mean line-of-sight velocity of Illustris subhalos. 
We started with all 7697 subhalos with more than $10^{4}$ stellar particles in the final output (redshift $z=0$) of the Illustris-1 run, identified in the SubFind Subhalo catalog \citep{vog14illintro,nel15illpub}. 
We refer to those subhalos simply as galaxies and to the set of 7697 galaxies as the global sample.

For each galaxy we constructed mean line-of-sight velocity maps of the stellar component in three projections with two different fields of view: $30\times30$\,kpc and $1.5\times1.5\,r_{\rm max}$, where $r_{\rm max}$ is the same radius as defined in \cite{illprol} as the radius at which the average 3D stellar density in a shell between $0.8r_{\rm max}$ and $1.25r_{\rm max}$ is equal to $1.4\times10^{4}$\,M$_{\sun}$\,kpc$^{-3}$. 
The typical surface brightness at the projected radius $r_{\rm max}$ of spherical galaxies is 29.5 (29.0) mag\,arcsec$^{-2}$ for the mass-to-light ratio of 5 (3).
The projection planes of the maps correspond to the principal planes of the respective galaxy inferred using the inertia tensor for stellar matter inside $r_{\rm max}$.

For each field of view, we constructed maps with fixed bin sizes as well as maps with Voronoi tessellation. 
The latter uses the Voronoi relaxation algorithm \citep{vor1,vor2} to divide the map into bins with similar stellar masses.
First, we bin the map into a $750\times750$ mesh and apply Gaussian smoothing with the radius equal to 0.27\,kpc, then we run the Voronoi relaxation algorithm.
We overlay the tessellated map with a white mask corresponding to those areas of the original $750\times750$ mesh where the surface density is below 96\,M$_{\sun}$\,kpc$^{-2}$ (the value was chosen, in part arbitrarily, in order to have the borders of the white area similar to the outer isophots in surface-brightness maps).
This kind of maps can be seen in Figs.~\ref{fig:ex}, \ref{fig:apsn}, \ref{fig:dispex}, and  \ref{fig:doubex} for the kinematics of the galaxies.
All our kinematic maps are mass-weighted.

We performed a visual inspection of the maps and selected 134 Illustris galaxies with a KDC visible in at least one projection plane. 
All selected galaxies were reviewed by two members of our team.
The galaxies in the Illustris-1 run have a limited resolution -- the gravitational softening length is 0.7\,kpc (1.4\,kpc) for the baryonic (dark-matter) component, thus sub-kpc kinematic features are not reliably resolved. 
For this reason our sample does not contain small, compact KDCs such as those reported in fast rotators by \cite{mcd06}. 
In order to have a sufficiently large sample, we included proper kinematically distinct cores as well as more extended central components that often exceed the effective radius of the host galaxy (for more details on the host effective radii, see Sect.\,\ref{sec:age} and Fig.\,\ref{fig:reff}). 
However, to include a galaxy in our sample, we require the KDC to cover at most roughly about half of the wider, $1.5\times1.5\,r_{\rm max}$, fields of view of the velocity maps.
The most extended KDC in our sample turns out to have the radius $r_{\rm KDC}=0.4r_{\rm max}$, while half of the sample has $r_{\rm KDC}<0.2r_{\rm max}$. 
We do not include rapidly rotating cores that have similar orientation of the kinematic axis as the outer parts of the galaxy. 
Fig.\,\ref{fig:ex} shows four examples of galaxies from our Illustris KDC sample. 
The galaxies were chosen to show examples of different KDC evolutions (see Sect.\,\ref{sec:ex}; properties of these four galaxies are listed in Tab.\,\ref{tab:ex}). 
Throughout the paper (except Fig.\,\ref{fig:dekin}) $x$, $y$, and $z$ denote the major, intermediate, and minor axis of a galaxy, respectively. 
We refer to galaxies by their SubFind ID from the last snapshot throughout their entire evolution, even though their progenitors in previous snapshots have different SubFind IDs.

We measure the KDC radii, $r_{\rm KDC}$, combining information about the stellar component from the velocity maps and two radial profiles: (1) the radial profile of the mean tangential velocity, $v_{\perp,\rm mean}$, in spherical bins in 3D and in circular bins for the projected data; (2) the radial profile of the angular difference between the directions of the angular momenta inside and outside of the given radius, $\alpha(L_{\rm in},L_{\rm out})$, for 3D and projected data, where $\alpha$ is defined as follows
\begin{equation}
\cos\alpha(L_{\rm in},L_{\rm out}) = \mathbf{\hat L}_{\rm in}\cdot\mathbf{\hat L}_{\rm out}.
\end{equation}
Fig.\,\ref{fig:cos12} shows the radial profiles for the four galaxies from Fig.\,\ref{fig:ex}.
The KDC radius inferred from the velocity maps corresponds well to the minimum or to the end of the decline of $v_{\perp,\rm mean}$.
It is also usually near the maximum value and/or near the position of maximum gradient of $\alpha(L_{\rm in},L_{\rm out})$, but the correlation is not so tight and the radial shape of $\alpha(L_{\rm in},L_{\rm out})$ depends on many factors such as the relative orientation of the kinematic components, the relative magnitude of their velocities, the size and shape of the KDC, etc. 
We made attempts to use $\alpha(L_{\rm in},L_{\rm out})$ to automatically identify KDCs in Illustris galaxies, but did not find much success.
Some KDC hosts have relatively small maximal values of $\alpha(L_{\rm in},L_{\rm out})$ and similar values are often produced in other galaxies due to other features (e.g., streams and tails in outer parts or kinematic twists) or due to inherent fluctuations.
Therefore, we resorted to the visual inspection of the velocity maps. 
The precision of the measurements of $r_{\rm KDC}$ is around 1\,kpc, 0.5\,kpc for the smallest KDC.

\begin{figure} [!htb]
\includegraphics[width=\hsize]{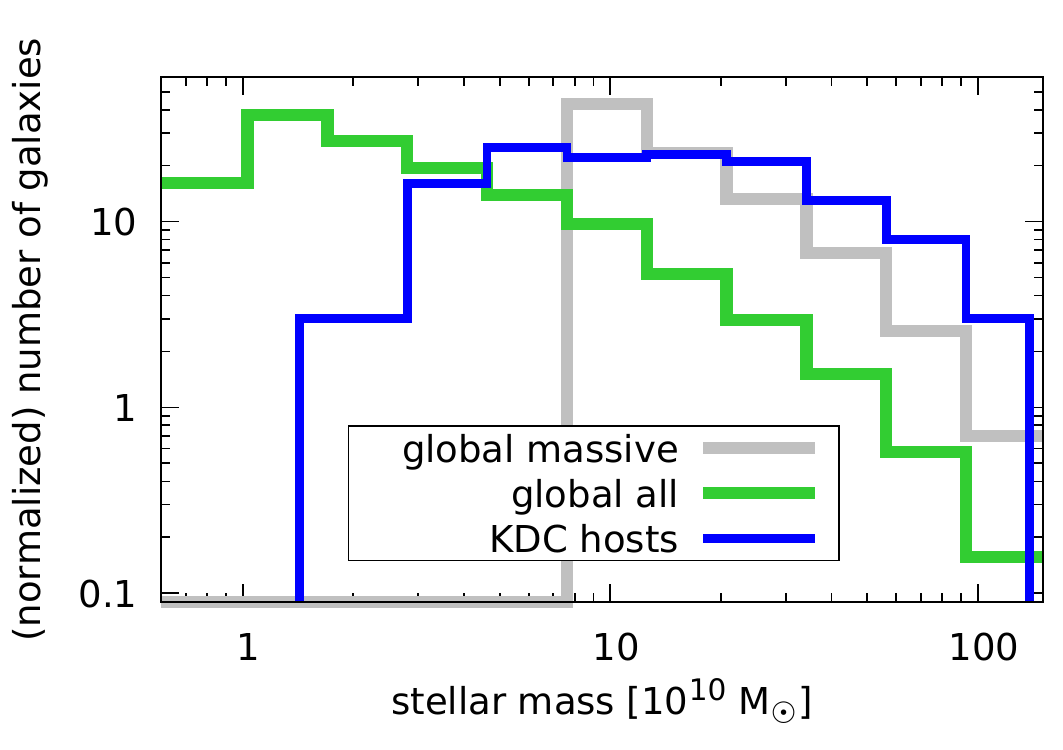}
\caption{
Comparison of the stellar mass distribution of the KDC hosts with the global sample. 
The distribution of the global sample is normalized to the same total number as for the KDC hosts.
The grey histogram shows the mass distribution of the global sample for galaxies more massive than $7.4\times10^{10}$\,M$_{\sun}$ normalized to the same number as the KDC hosts in the mass range.
\label{fig:Mdist}}
\end{figure}

Fig.\,\ref{fig:Mdist} compares the distribution of stellar mass inside $r_{\rm max}$ at redshift $z=0$ for the KDC hosting galaxies with the global sample of 7697 galaxies. 
The global sample is normalized to the KDC host sample. 
The distribution of the KDC hosts is truncated for low-mass galaxies probably because the resolution of these galaxies in the simulation is not sufficient for us to detect KDCs. 
Thus, our KDC sample is incomplete for low-mass galaxies.
The grey histogram in Fig.\,\ref{fig:Mdist} shows the mass distribution for the 1160 galaxies with masses higher than $7.4\times10^{10}$\,M$_{\sun}$ normalized to the 90 KDC hosts in the same mass range. 
The threshold $7.4\times10^{10}$\,M$_{\sun}$ is the lower boundary of the first bin where the detection of the KDCs is reliable since the galaxies are well resolved and there is no longer a significant increase of detected KDCs with the mass of the galaxy. 
For the well-resolved galaxies, we see that KDC hosts tend to follow the global mass distribution with a slight preference towards massive galaxies. 
This is interesting in comparison with Illustris galaxies with prolate rotation that are clearly biased towards more massive galaxies, see Figure\,4 in \cite{illprol}.

\begin{figure} 
\includegraphics[width=\hsize]{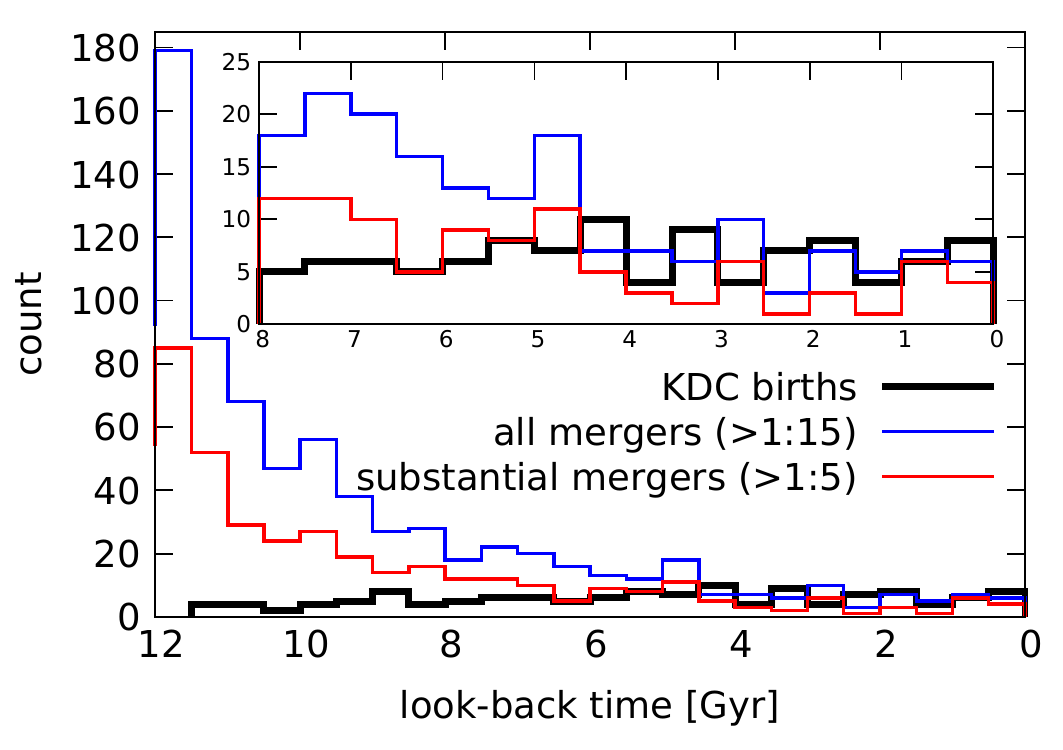}
\caption{
Histograms of time distribution of KDC births and mergers undergone by their host galaxies.
The inset shows zoom on the last 8\,Gyr.
\label{fig:merbir}}
\end{figure}

\subsection{Times of the births of KDCs} \label{sec:birth}

To determine the times in which the KDCs emerged, we used Illustris data from snapshots 60\,--\,135 corresponding to last 11.6\,Gyr of evolution (redshifts $3.0>z>0$).
The look-back times are computed with the cosmological parameters consistent with the \textit{Wilkinson Microwave Anisotropy Probe}-9 measurements \citep{wmap9} as these parameters are also adopted in the Illustris simulation. 
The average value of the time spacing between snapshots is 156\,Myr during the last 10\,Gyr of the simulation.

We followed the main progenitor branch of the Illustris SubLink merger trees \citep{rg15illmer} for all of our 134 KDC hosts. 
We plotted the maps of stellar velocity as well as the radial profiles of  $v_{\perp,\rm mean}$ and $\alpha(L_{\rm in},L_{\rm out})$ (see Fig.\,\ref{fig:cos12}) for all 76 snapshots for each galaxy. 
The maps were plotted in the three projection planes that correspond to the principal planes in the last snapshot. 
We kept the orientation of the planes in all snapshots the same even though they are not the principal planes of the galaxy in the earlier snapshots. 

We inspected simultaneously the maps and the radial profiles for the same snapshots. 
We mark the earliest snapshot at which we can see the KDC. 
Merger events make the plots chaotic, thus if we were not able to see the KDC for a few snapshots, we still counted this as a continuous existence of the KDC. 
We allow the KDC to not be visible in approximately three snapshots -- the number is not exact since, in such cases, one or more snapshots usually have borderline visibility of the KDC.
In case the KDC is not evident for a larger number of snapshots, we marked the subsequent emergence of the KDC as the KDC birth. 

The black histogram in Fig.\,\ref{fig:merbir} shows the distribution of look-back times of the KDC births. 
The distribution is almost uniform with a slight deficiency in birth times larger than 9 Gyr. 
This is in contrast to the prolate rotation in Illustris, where almost all cases were younger than 6\,Gyr \citep{illprol}.

\begin{figure} 
\includegraphics[width=\hsize]{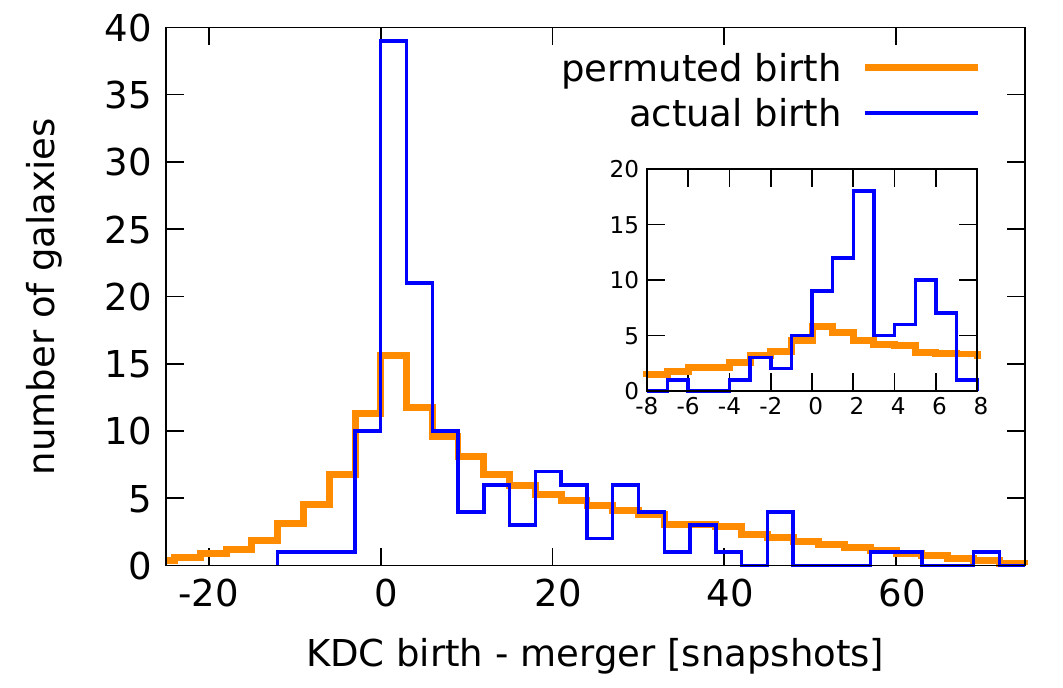}
\caption{
Distribution of the snapshot difference between the KDC birth and the closest merger undergone by the host galaxy for the actual KDC births (in blue) compared to the distribution where the KDC births were permuted within the sample (in orange), with the bin size of 3 snapshots.
The inset is a zoom around zero for the actual births, with the bin size of one snapshot. 
One snapshot is approximately 0.16\,Gyr.
\label{fig:perm}}
\end{figure}

\subsection{Mergers} \label{sec:merger}

We analyzed the merger history of our KDC hosts using the SubLink trees. 
We followed the branches of the next progenitors (secondaries) that 
(1) occurred in the merger tree of the primary galaxy during the last 12.6\,Gyr (since snapshot number 49, redshift $z=5.0$)\footnote{We note that snapshots 53 and 55 of Illustris-1 run are not available.}; 
(2) had at least 5\,\% baryonic mass of the primary in at least 4 snapshots during the last 20 snapshots of their existence. 
There are 1677 such secondaries for our sample of 134 KDC hosts.
For these secondaries, we compute the merger mass ratio as the ratio of stellar matter of the merger progenitors in the snapshot in which the secondary progenitor reaches its maximum stellar mass, as suggested in \cite{rg15illmer}, but only in the last 20 snapshots of their existence.
For secondaries that had more stellar mass than the primary in the snapshot of the mass measurements, the merger mass ratio is set to 1:1.
We mark the time of the merger at the look-back time of the last snapshot in which the secondary progenitor still has at least 10 stellar particles. 

During all these calculations, we prohibit the secondary galaxy from gaining stellar particles from the primary. 
That means that we check all IDs of the stellar particles in the secondary and if there are particles that belonged to the primary in previous snapshots, these particles are treated as a part of the primary.
This action is motivated by the structure of the Illustris catalog.
The SubFind algorithm ran independently on each snapshot data. 
In case of a merger of two similarly massive galaxies, often the majority of particles is assigned to the primary and secondary progenitors alternately in subsequent snapshots.
This would make the measurements of the merger mass ratio incorrect.

From now on, we consider only the mergers with the stellar-mass ratio at least 1:15 and the time of the merger younger than 12.2\,Gyr (snapshot number 54, redshift $z=4.0$).
One KDC host had no such merger. 
The remaining 133 KDC hosts collectively underwent exactly 800 mergers, between 1 and 14 mergers for an individual host. 
More than half of these mergers took place more than 10\,Gyr ago, as shown in the blue histogram in Fig.\,\ref{fig:merbir}, while only 10 KDCs are that old. 
The red histogram in Fig.\,\ref{fig:merbir} shows the distribution of 418 mergers with the stellar-mass ratio at least 1:5 -- we refer to them as \textit{substantial mergers}.

From the 800 mergers, we selected \textit{closest mergers} to the KDC birth (Sect.\,\ref{sec:birth}) for the 133 hosts that experienced a merger. 
We picked the merger with the smallest snapshot difference. In case there are more mergers with the same difference, we took the one with the highest merger mass ratio.
The distribution of the snapshot difference is shown as the blue histogram in Fig.\,\ref{fig:perm}.
It shows a strong peak around zero.
To prove that the peak is significant, we repeated the selection of the closest mergers, but with the snapshot numbers of the KDC birth permuted within the sample of 134 KDC hosts -- the orange histogram in Fig.\,\ref{fig:perm}.

The permuted sample is averaged over all possible permutations. 
We did it by pairing each KDC birth with all other 133 merger histories within the sample and we took each pair with the weight of $1/133$.
This way, we compare the merger snapshots with a randomized KDC birth for each galaxy, but we are able to keep the same distribution of KDC births.
The histogram for the permuted births indeed shows much less pronounced peak of the closest mergers. 
Both distributions, for permuted and actual births, have a long tail in positive values just because a galaxy is more likely to experience mergers at early times, see Fig.\,\ref{fig:merbir}.

The inset in Fig.\,\ref{fig:perm} shows the snapshot difference around zero with the bin size of one snapshot.
In this histogram, we see that the highest peak of the actual births is, in fact, at the snapshot difference equal to two.
That means that the KDC is most likely to emerge two snapshots (approximately 0.3\,Gyr) after the merger. 
Following this finding, we consider a merger to be associated with the KDC birth, if the merger occurs $\pm6$ snapshots (within $\pm1$\,Gyr) from the peak of the distribution, i.e. the KDC birth occurred from four snapshots before to eight snapshots after the merger. 
There are 81 KDC hosts with 125 mergers in this asymmetric 13-snapshot window around the KDC birth, out of which 61 had only one merger, 12 had two, four had three, and the remaining four had 4, 5, 7, and 12 mergers.
For the cases with multiple mergers, we selected one \textit{associated merger} with the highest merger mass ratio.

\begin{figure} [!htb]
\includegraphics[width=\hsize]{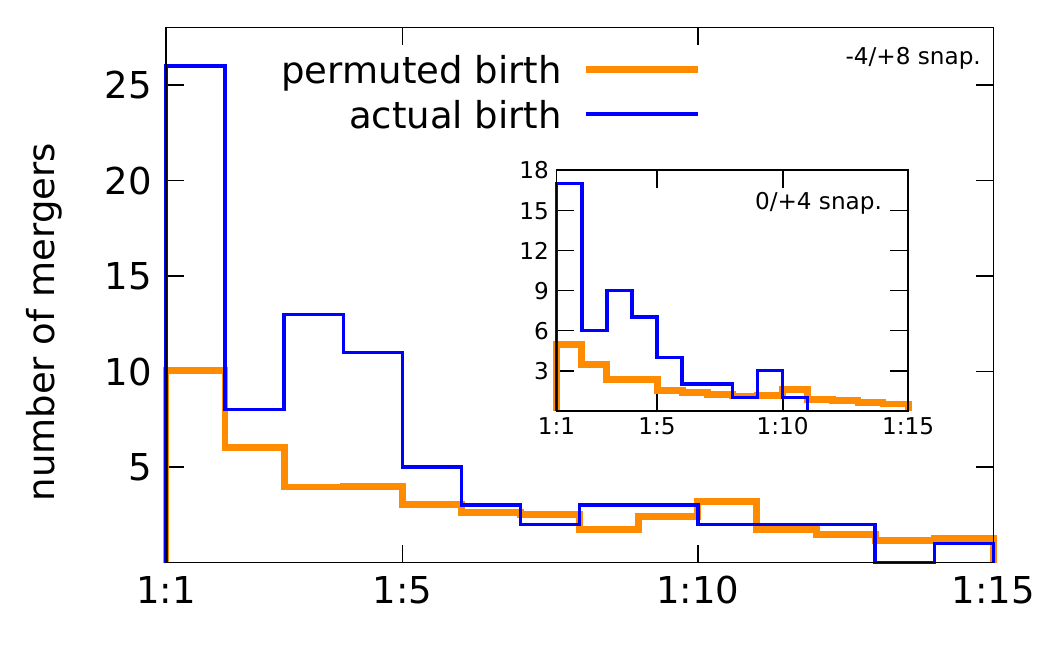}
\includegraphics[width=\hsize]{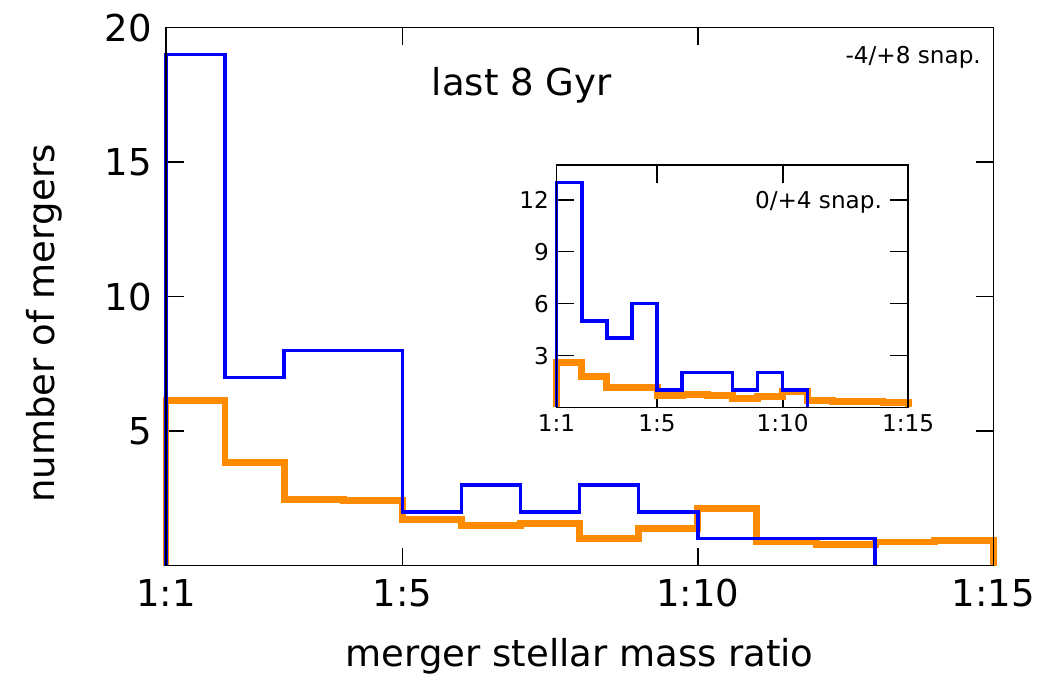}
\caption{
Distributions of the stellar-mass ratios for associated mergers for the actual KDC births (in blue) and permuted KDC births (in orange). 
The upper panel concerns the whole investigated period of 12.2\,Gyr, the lower panel shows the associated mergers for the last 8\,Gyr of the simulation.
The insets show the same distributions for tightly associated mergers.
\label{fig:mrat}}
\end{figure}

In Fig.\,\ref{fig:mrat}, we compare the distribution of the merger stellar-mass ratios of the associated mergers for the actual and permuted KDC births. 
The insets show the same comparison, but for more \textit{tightly associated mergers}, where the KDC birth occurred between the merger snapshot to four snapshots after the merger -- mergers within an asymmetric 5-snapshot window around the KDC birth.
The histograms show that the association between the actual births and mergers is significantly higher than for the `randomized' permuted sample, mainly for the substantial mergers.
From all 134 KDCs and all 800 mergers during the whole investigated period of the last 12.2\,Gyr of the simulation, 81 (52) KDCs with the actual births and 45.1 (23.8) KDCs with the permuted births had (tightly) associated mergers and are displayed in the upper panel of Fig.\,\ref{fig:mrat}.

\begin{figure} [!htb]
\includegraphics[width=\hsize]{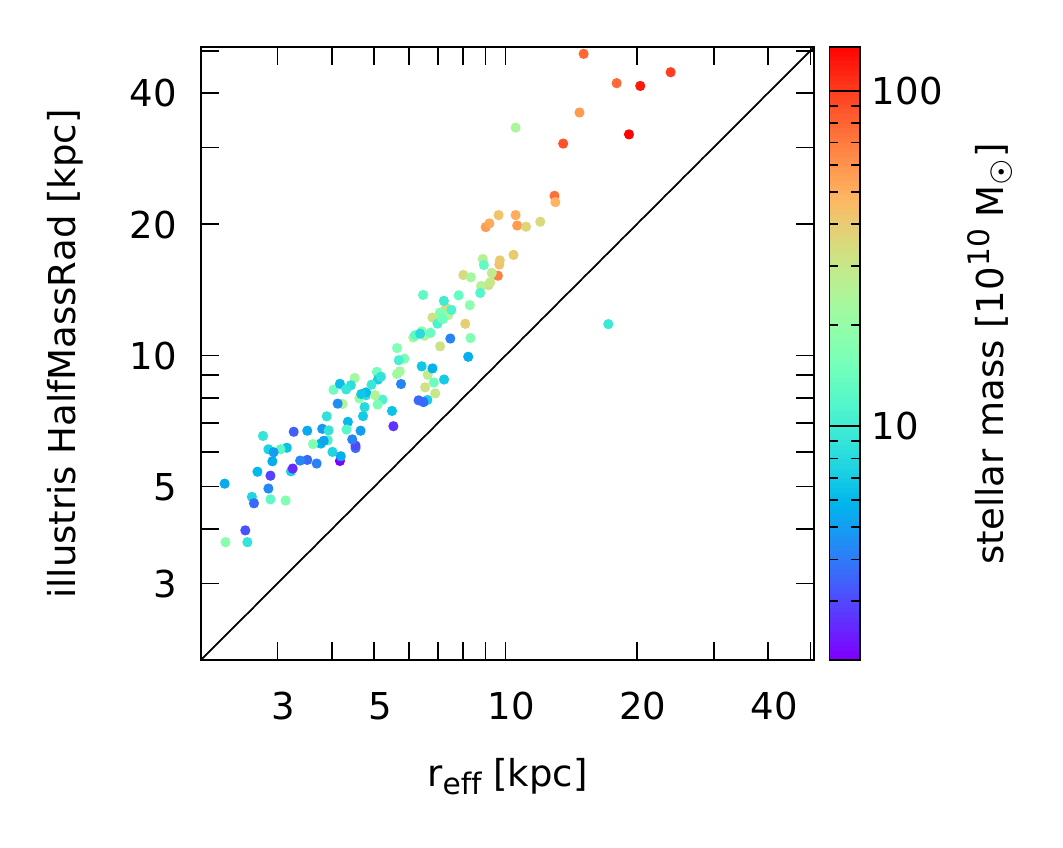}
\caption{
Correlation of the average of effective radii from de Vaucouleurs' fits ($r_{\rm eff}$)  in the three principal projection planes of the respective galaxies with the half mass radius for stellar particles given by the Illustris subhalo catalog (HalfMassRad) for our KDC host sample at $z=0$.
Colors code the stellar mass of the host galaxies.
\label{fig:reff}}
\end{figure}

Since there is an abundance of mergers at higher look-back times, the probability of the merger to be associated by coincidence is higher.
Therefore, we checked the associated mergers for a subset of 103 KDCs that emerged during the last 8\,Gyr of the simulation, where the mergers are sparser.
Overall, the 103 hosts with the actual KDC births underwent 148 mergers and the 103 hosts with the permuted birth 134 mergers during this period, 
out of which 57 (37) KDCs with the actual births and 27.6 (12.3) KDCs with the permuted births had (tightly) associated mergers.
The distribution of their mass ratios is shown in the lower panel of Fig.\,\ref{fig:mrat}.

As expected, the percentage of KDC hosts with associated mergers decreased slightly in the period with less frequent mergers -- 60\,\% for the whole period and 55\,\% for the last 8\,Gyrs. 
However, the mergers during the shorter and more recent period are more likely to be associated with the KDC birth -- 16\,\% of the mergers fall into the 13-snapshot window around the KDC birth during the whole period, while for the last 8\,Gyr it is 46\,\%.
Overall, the KDC birth is more likely to be associated with major mergers and, in the same period, major mergers seem to be more tightly associated with the KDC birth. 
From the 81 associated mergers, 56 are substantial (the stellar-mass ratio at least 1:5) and only 7 have the stellar-mass ratio between 1:10 and 1:15.

\subsection{KDC stellar ages and sizes} \label{sec:age}

For each galaxy in our sample, we adopted an effective radius, $r_{\rm eff}$, as an average of effective radii from de Vaucouleurs' fits in the three principal projection planes. 
The fits  are performed on the average surface density of stellar particles in circular bins between 1\,kpc and $r_{\rm max}$. 
The average $r_{\rm eff}$ corresponds very well to the effective radius in $xz$-plane. 
Fig.\,\ref{fig:reff} shows the correlation of $r_{\rm eff}$ and the half mass radius for stellar particles given by the Illustris subhalo catalog for our sample at $z=0$. 
Especially for massive galaxies, the Illustris radius is larger probably due to the richness of the tidal structures in the outer parts of these galaxies and because the outer parts of nearby smaller galaxies are usually assigned to the massive galaxy by the SubFind algorithm. 
We believe our approach to be a good approximation to the effective radius that would be inferred observationally.

\begin{figure*} [!htb]
\centering
\includegraphics[width=\hsize]{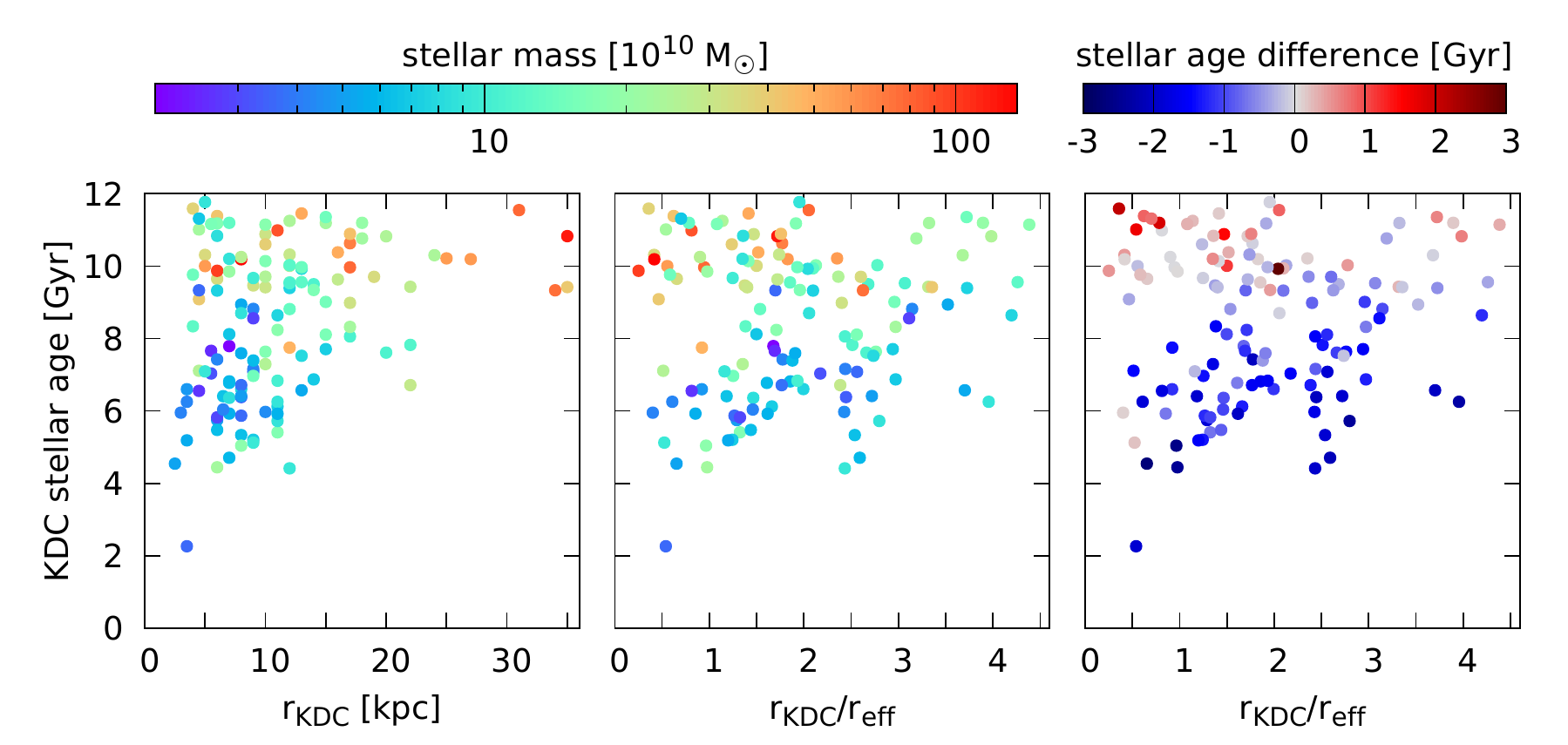}
\caption{
KDC sizes versus the mass-weighted stellar ages inside $r_{\rm KDC}$.
First panel shows the actual physical sizes of the KDCs, the other two panels show their relative size $r_{\rm KDC}/r_{\rm eff}$.  
Colors code the stellar mass of the host galaxies at redshift $z=0$ for the first two panels and the age difference between stars inside and outside $r_{\rm KDC}$ (inner minus outer) for the last panel.
\label{fig:sa}
}
\end{figure*}

Fig.\,\ref{fig:sa} shows KDC sizes, and KDC stellar ages, and host masses, all measured in the last snapshot of the simulation at $z=0$.
The KDC stellar age is measured as the mean mass-weighted age of stellar particles inside the KDC radius $r_{\rm KDC}$. 
The first panel shows the actual physical sizes of the KDCs versus their stellar ages. 
The vertical alignment of the points is due to the limited resolution of $r_{\rm KDC}$.
Even though our KDC sample covers a different range of KDC sizes, the graph partially recovers the trend reported for the SAURON sample \citep[][and their Figure\,16; see also Sect.\,\ref{sec:intro}]{mcd06} -- KDCs with older stellar ages cover a wide range of sizes, while younger stars are associated with smaller values of $r_{\rm KDC}$.
In our case, this trend is mostly driven by the link between the KDC size and the size (or mass) of the host galaxy. 
In the middle panel of Fig.\,\ref{fig:sa} the KDC radius is replaced with the KDC relative size $r_{\rm KDC}/r_{\rm eff}$ -- the ratio of the KDC radius and the host galaxy effective radius, and the above-described trends mostly disappear.

Color codes the host stellar mass in the first two panels.
In the last panel of Fig.\,\ref{fig:sa}, the color reflects the difference between the mean mass-weighted age of stars inside and outside $r_{\rm KDC}$.
The outer region includes stellar particles between the $1.1r_{\rm KDC}$ and the radius $r_{\rm max}$ as defined in Sect.\,\ref{sec:selection}.
In the case that $0.1r_{\rm KDC}$ is smaller than 0.7\,kpc (the gravitational softening length), the inner radius is set to $r_{\rm KDC}+0.7$\,kpc.
The age difference is mostly determined by the KDC stellar age and ranges from -2.7\,Gyr for the young KDC stars to 3.0\,Gyr for the old ones. 
Most galaxies with close-to-equal inner and outer stellar ages occur around the age 10\,Gyr.

\begin{figure*} [!htb]
\centering
\includegraphics[width=0.7\hsize]{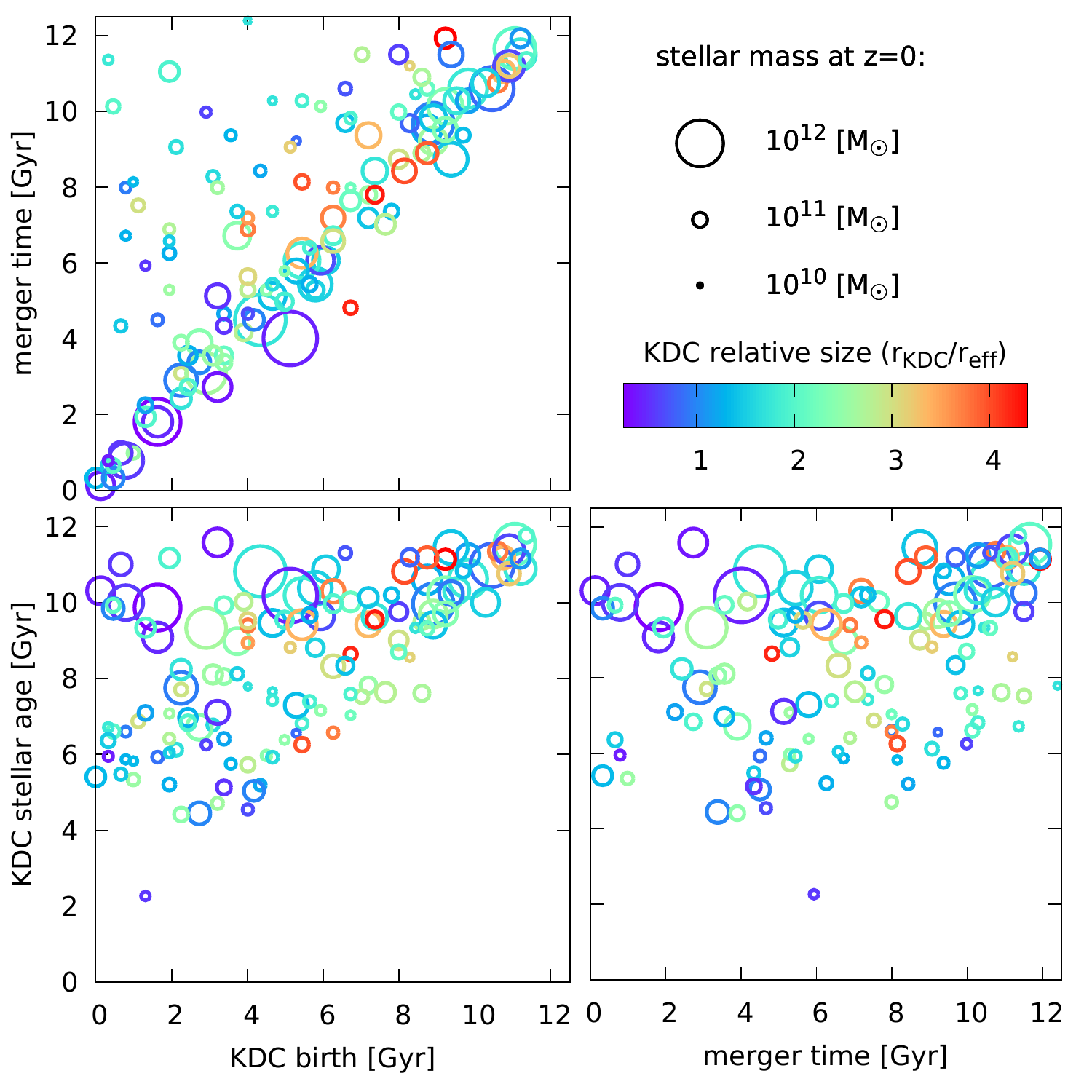}
\caption{
Correlations between the time of the KDC birth and the associated or closest merger and the mass-weighted stellar age inside $r_{\rm KDC}$.
The time labels are expressed in terms of look-back time.
Circle areas are proportional to the stellar mass of the galaxies at redshift $z=0$. 
Colors reflect the relative sizes of the KDCs at $z=0$.
\label{fig:agebir}
}
\end{figure*}

\subsection{Birth-merger-age correlations} \label{sec:corr}

Fig.\,\ref{fig:agebir} shows correlations between the KDC birth, mergers and the KDC stellar age.
For the 81 KDCs that have an associated merger, we display the associated merger, for the remaining KDCs, 52 are paired with the closest merger and the one KDC host that did not experience any merger in the probed period, is paired with its last merger 12.39\,Gyr ago, even though the merger occurred just outside the probed period (see Sect.\,\ref{sec:merger}).
The color codes the KDC relative size $r_{\rm KDC}/r_{\rm eff}$ -- the ratio of the KDC radius and the host galaxy effective radius (see Sect.\,\ref{sec:age}). 
The size of the circles is proportional to the stellar mass of the galaxies at the end of the simulation.
Note that the inclination of massive galaxies towards old stellar ages holds generally for all the galaxies, with or without the KDC. 

The tightest correlation is between the KDC births and mergers (the top panel of Fig.\,\ref{fig:agebir}). 
The correlation is especially strong for the massive galaxies -- 63 out of 75 (84\,\%) hosts with the stellar mass larger than $10^{11}$\,M$_{\sun}$ have an associated merger, while for the 59 less massive hosts it is only 18 (30\,\%).
Even though some galaxies experienced more mergers after the KDC birth, almost all the hosts without the associated mergers underwent the closest merger prior to the KDC birth.
 
All the KDC stellar ages are close to the KDC birth or older (the bottom left panel of Fig.\,\ref{fig:agebir}), meaning that inside the KDC radius, there are not many stars significantly younger than the KDC birth. 
However the KDC stellar ages can be significantly larger than the time of the KDC birth. 
The correlation between the KDC stellar ages and birth seems to be independent of the associated mergers.
For both groups -- KDCs with and without associated mergers, a quarter of the KDC stellar ages is within 1\,Gyr from the birth (21 out of 81 with the associated merger and 13 out of 53 without one).

The KDC stellar ages and mergers show no apparent correlation (the right panel of Fig.\,\ref{fig:agebir}).
This is a result of the above-mentioned findings: 
(1) some closest mergers are not associated with the KDC birth,  
(2) there are a lot of KDCs with stellar ages older than KDC birth, and
(3) the age-birth correlation is partially independent of the birth-merger. 

There seems to be no clear correlation between the KDC relative size (the circle colors in Fig.\,\ref{fig:agebir}) and the other quantities. 
The less compact KDCs tend to occur in less massive hosts, but this could be an effect of the selection procedure and the limited resolution of the low-mass galaxies. 
Nonetheless, it is worth noting that while compact KDCs are associated with the whole range of birth and merger times, the less compact KDCs tend to have earlier births and mergers (the left panels of Fig.\,\ref{fig:agebir}).
All 18 KDCs with $r_{\rm KDC}/r_{\rm eff}>3$ are born more than 4\,Gyr ago. 
These 18 KDCs have a slightly lower rate of the associated mergers (8 out of 18) than average, but they do not seem to be exceptional in any way.
In one to three cases, the KDC experienced some size growth. 
However, more often they are born large, in some cases even too large to satisfy our criteria for KDCs, but their size or relative size gets smaller over time. 

\begin{figure*} [!htb]
\centering
\includegraphics[width=\hsize]{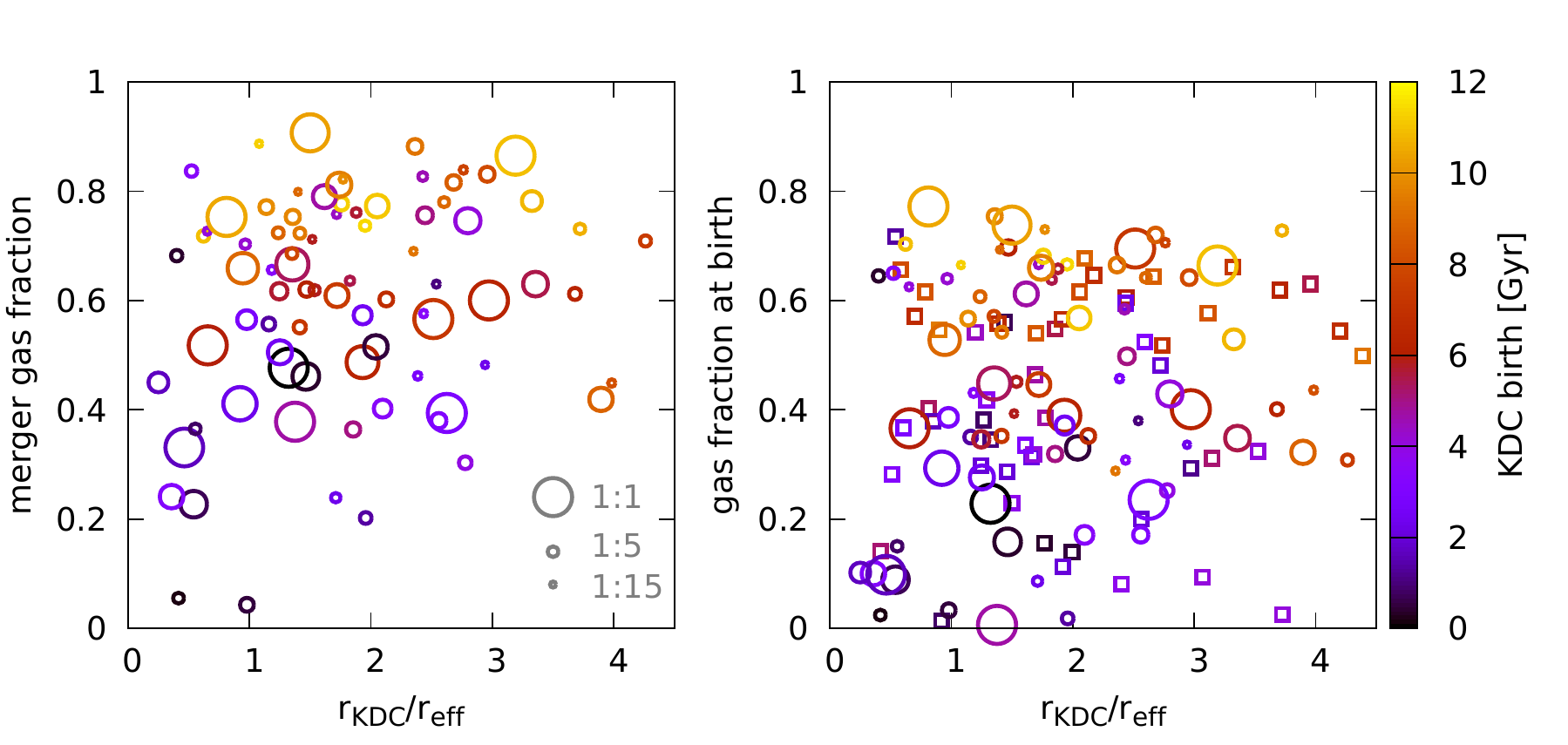}
\caption{
Left panel: KDC relative size $r_{\rm KDC}/r_{\rm eff}$ versus the total merger gas fraction for 81 KDCs with associated mergers.
Right panel: KDC relative size versus the total gas fraction at the time of the KDC births for all 134 KDCs.
In both panels, the color codes the look-back time of the KDC birth and, for the KDCs with associated mergers, the circle sizes are proportional to the merger stellar-mass ratio. 
KDCs without associated mergers are represented by the square symbols.
\label{fig:gas}
}
\end{figure*}

In addition, the KDC relative size seems not to be correlated with either merger mass ratio or the amount of gas present during the KDC birth, as shown in Fig.\,\ref{fig:gas}.
Horizontal axes of both panels show the KDC relative size and the symbol colors code the time of the KDC birth. 
The left panel shows the relative size versus the merger gas fraction for the 81 KDCs with associated mergers (see Sect.\,\ref{sec:merger}). 
The gas fraction is measured at the same time as the merger stellar-mass ratio (see Sect.\,\ref{sec:merger}) as the sum of the gas mass of both merger progenitors divided by the sum of the baryonic masses. 
The circle sizes are proportional to the merger stellar-mass ratio. 
The gas fraction corresponds well to birth time as the galaxies in the early universe had generally higher gas fraction.

The right panel of Fig.\,\ref{fig:gas} shows the relative size versus the gas fraction of the KDC host at the time of the KDC birth. 
KDCs with associated mergers are represented with the same symbols as in the left panel, KDCs without associated mergers with the squares. 
The host gas fraction at the KDC birth is lower than the merger gas fraction for 76 out of 81 KDCs with associated mergers.
On average, there is 67\,\% of the initial gas amount left in the system after the merger.
None of the quantities show a clear trend with the KDC relative size.

\begin{table*}
\caption{Properties of four selected Ilustris galaxies with KDCs}
\label{tab:ex}
\centering
\begin{tabular}{cccccccccccccccccccccccccccccc} 
\hline\hline
(1) & (2) & (3) & (4) & (5) & (6) & (7) & (8) & (9) & (10) & (11) & (12) & (13) & (14)\\
\# & $r_{\rm eff}$ & $r_{\rm max}$ & $M(r_{\rm max})$ & $f_{\rm g0}$ & $r_{\rm KDC}$ & $r_{\rm KDC}/r_{\rm eff}$ & Age(in) & Age(out) & Birth & $f_{\rm g,b}$ & Merger & Ratio & $f_{\rm g,m}$\\
 & (kpc) & (kpc) & ($10^{10}$\,M$_{\sun}$) & & (kpc) &  & (Gyr) & (Gyr) & (Gyr) &  & (Gyr) &  & \\
\hline
140593 & 9.1 & 77.8 & 33.3 & 0.01 & 6 & 0.66 & 9.6 & 9.5 & 5.93 & 0.37 & 6.07 & 1:1.0 & 0.52\\
353280 & 8.3 & 66.7 & 18.3 & 0.07 & 8 & 0.97 & 5.0 & 7.6 & 4.17 & 0.64 & 4.50 & 1:5.9 & 0.70\\
394799 & 4.7 & 40.5 & 6.6 & 0.04 & 7 & 1.50 & 8.1 & 9.1 & 3.73 & 0.23 & - & - & -\\
354838 & 2.9 & 49.5 & 12.4 & 0.00 & 4 & 1.38 & 8.3 & 9.9 & 6.58 & 0.56 & - & - & -\\
\hline
\end{tabular}
\tablefoot{
(1)	 the SubFind ID of the galaxy in the last snapshot of the Illustris-1 run; 
(2)	 $r_{\rm eff}$, the effective radius derived as an average of effective radii from de Vaucouleurs' fits in the three principal planes of the respective galaxy; 
(3)	 $r_{\rm max}$, the radius approximately corresponding to the surface density of 29 mag\,arcsec$^{-2}$, see Sect.\,\ref{sec:selection}; 
(4)	 $M(r_{\rm max})$, the stellar mass inside $r_{\rm max}$ at $z=0$; 
(5)	 $f_{\rm g0}$, the gas fraction at $z=0$ inside $r_{\rm max}$; 
(6)	 $r_{\rm KDC}$, the KDC radius; 
(7)	 $r_{\rm KDC}/r_{\rm eff}$, the KDC relative size; 
(8)	the mean mass-weighted age of stellar particles inside the KDC radius $r_{\rm KDC}$; 
(9)	the mean mass-weighted age of stellar particles between $r_{\rm KDC}$ and $r_{\rm max}$; 
(10)	the look-back time of the KDC birth; 
(11)	 $f_{\rm g,b}$, the total gas fraction at the time of the KDC birth; 
(12)	the look-back time of the associated merger (if any), see Sect.\,\ref{sec:merger};  
(13)	the stellar-mass ratio of the associated merger; 
(14)	 $f_{\rm g,m}$, the total gas fraction of the associated merger. 
}
\end{table*}

\begin{figure*} [!htb]
\resizebox{\hsize}{!}{\includegraphics{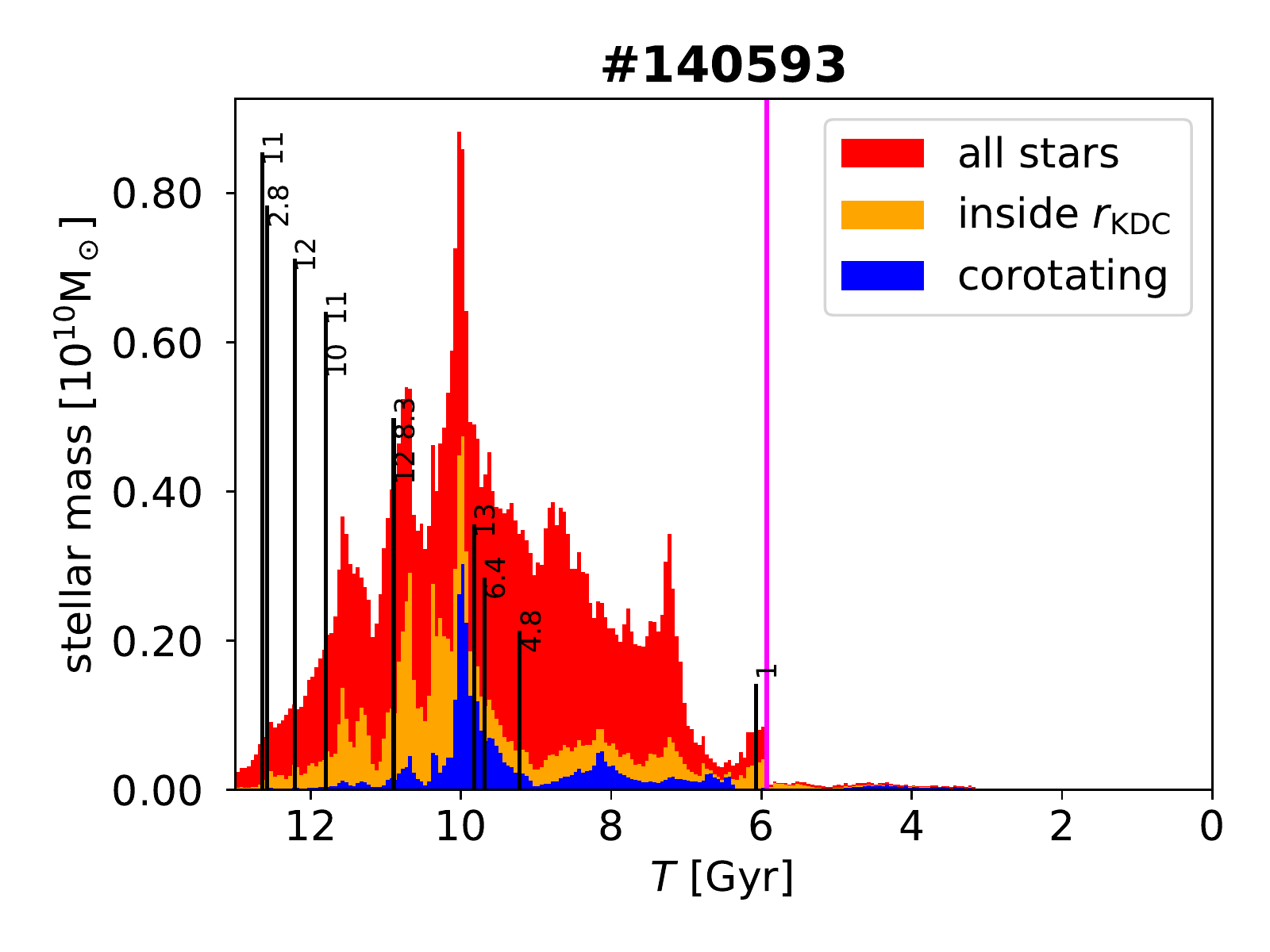}\includegraphics{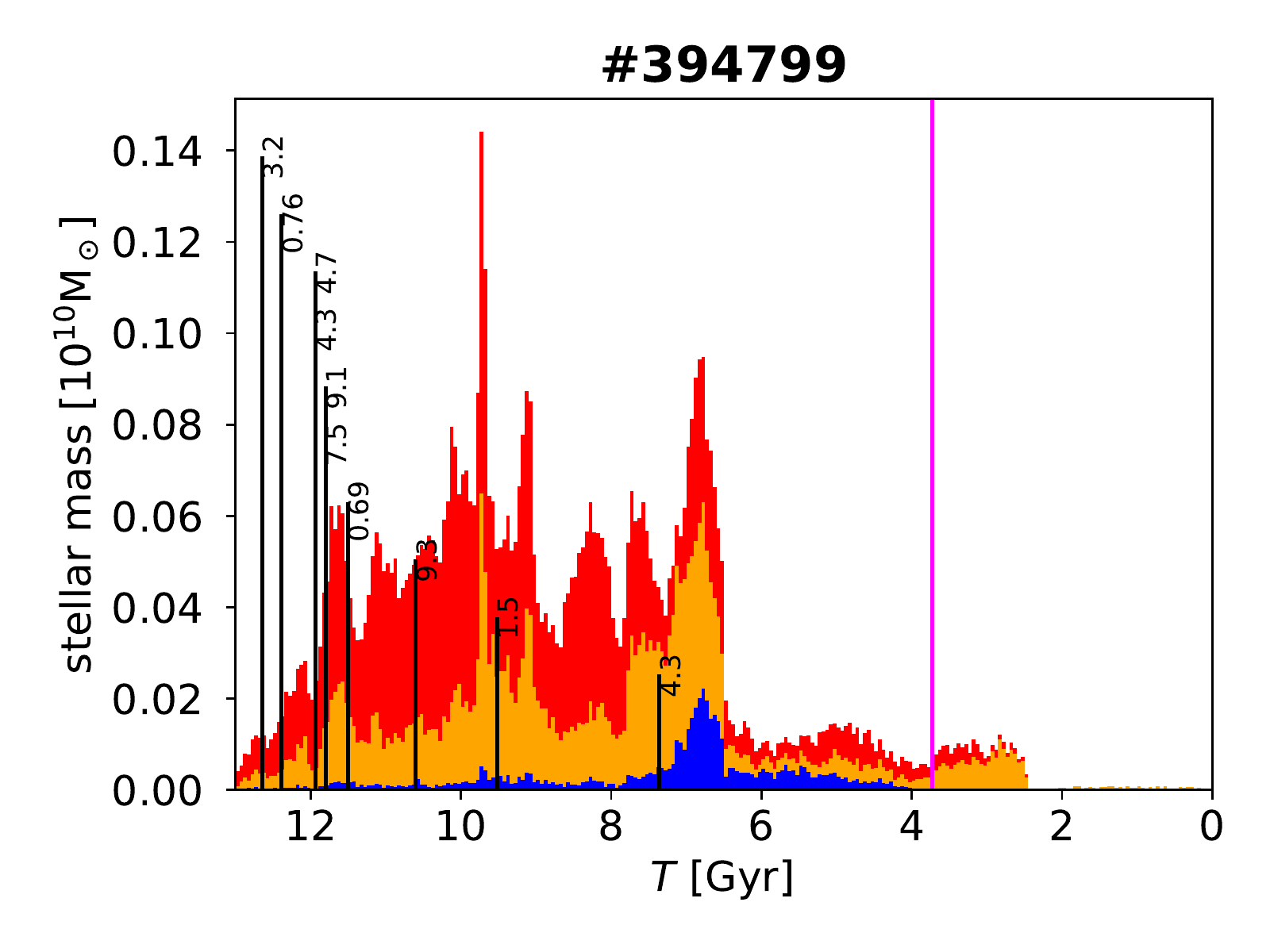}}
\resizebox{\hsize}{!}{\includegraphics{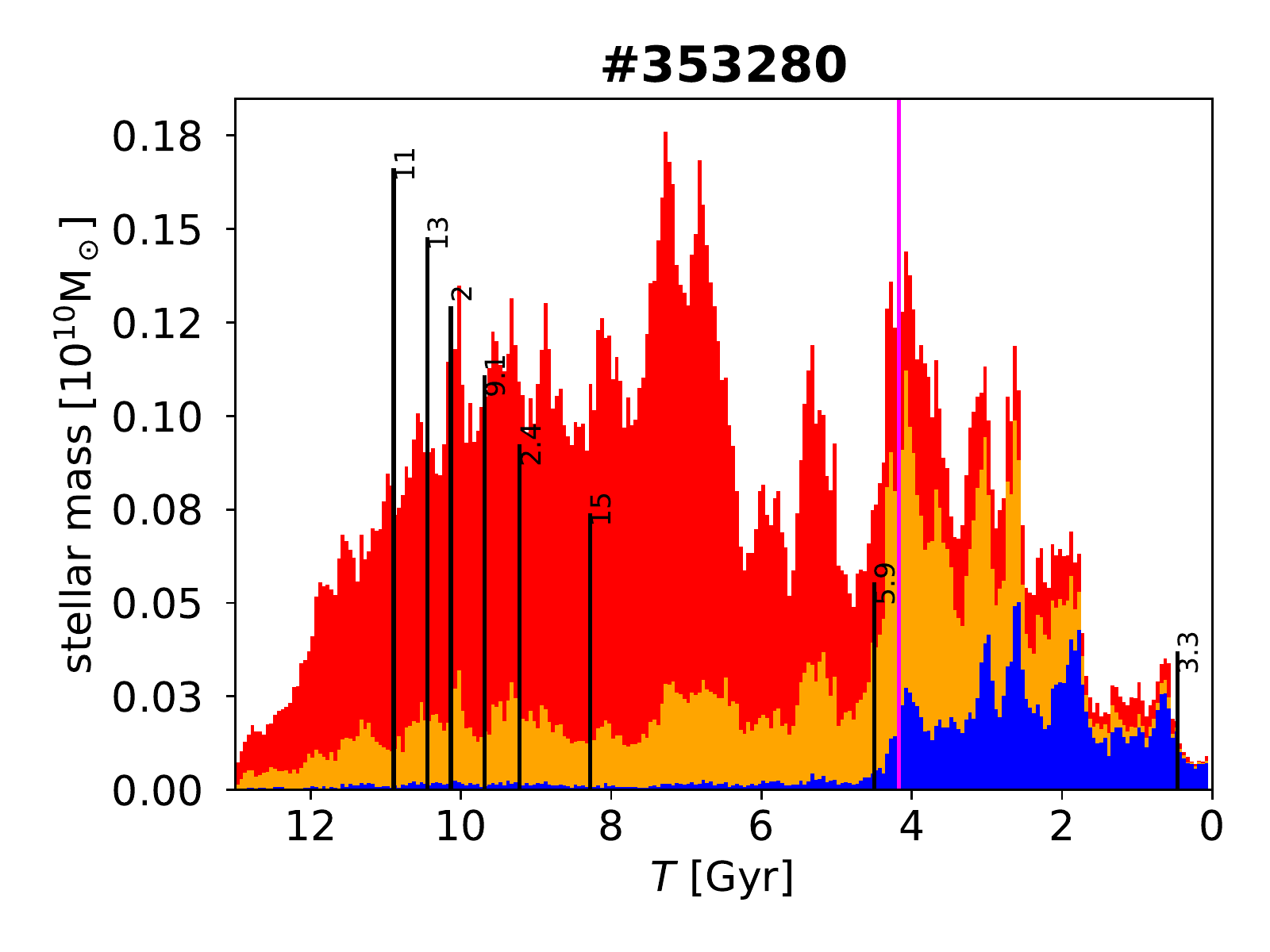}\includegraphics{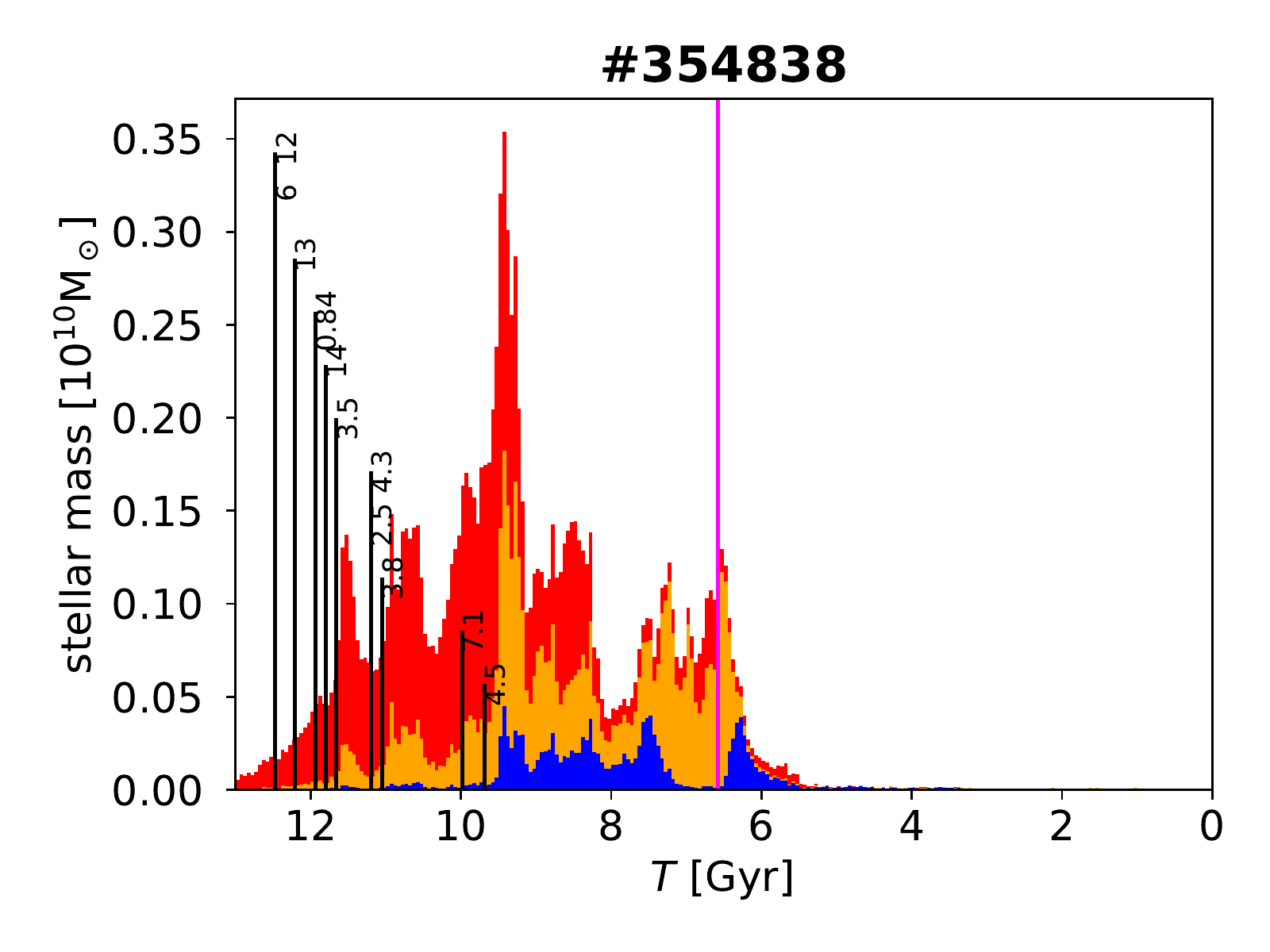}}
\caption{
Histograms of stellar ages of the four examples of Illustris galaxies with KDCs 
(properties of the galaxies are listed in Tab.\,\ref{tab:ex}).
The black and magenta vertical lines indicate the look-back time of the mergers and KDC birth, respectively.
The numbers indicate stellar-mass ratios of the mergers, e.g. ‘13’ means a merger 1:13.
\label{fig:shist}}
\end{figure*}

\subsection{Examples of KDC origins} \label{sec:ex}

In this section, we show four examples of the KDCs from our sample in more detail.
We take the same galaxies as displayed in Fig.\,\ref{fig:ex} (maps of surface density and line-of-sight kinematics) and Fig.\,\ref{fig:cos12} (radial profiles of $v_{\perp,\rm mean}$ and $\alpha(L_{\rm in},L_{\rm out})$). 
Properties of these galaxies and their KDCs are listed in Tab.\,\ref{tab:ex}.
Two of these galaxies (\#140593 and \#353280) had an associated merger, the other two (\#394799 and \#354838) had the last merger (bigger than 1:15) several gigayears before their KDC emerged.

Fig.\,\ref{fig:shist} shows histograms of stellar ages for all four galaxies. 
The magenta vertical lines indicate the KDC birth.
The black lines denote all the mergers undergone by the galaxy (see Sect.\,\ref{sec:merger}). 
The numbers indicate stellar-mass ratios of the mergers, e.g. ‘13’ means a merger 1:13.
The histograms are computed from the ages of the stellar particles present in the galaxy in the last snapshot. 
The red part corresponds to all stellar particles, yellow to the particles inside $r_{\rm KDC}$. 
Inside $r_{\rm KDC}$, there is 29, 35, 43, 45\,\% of the total stellar matter for the galaxy \#140593, \#353280, \#394799, and \#354838, respectively. 
The blue histograms include particles inside $r_{\rm KDC}$ that contribute to KDC rotation.
They are selected iteratively in the following manner: the angular momentum $L_{\rm KDC}$ of particles is calculated and only particles with angular momentum vectors within $\pi/6$ from the direction of $L_{\rm KDC}$ are retained. 
The procedure starts with all particles inside $r_{\rm KDC}$ and continues until convergence.
Particles included in the blue histograms represent 33, 28, 17, 24\,\% of the stellar matter inside $r_{\rm KDC}$ for the galaxy \#140593, \#353280, \#394799, \#354838, respectively.

\begin{figure*} [!htb]
\centering
\includegraphics[width=\hsize]{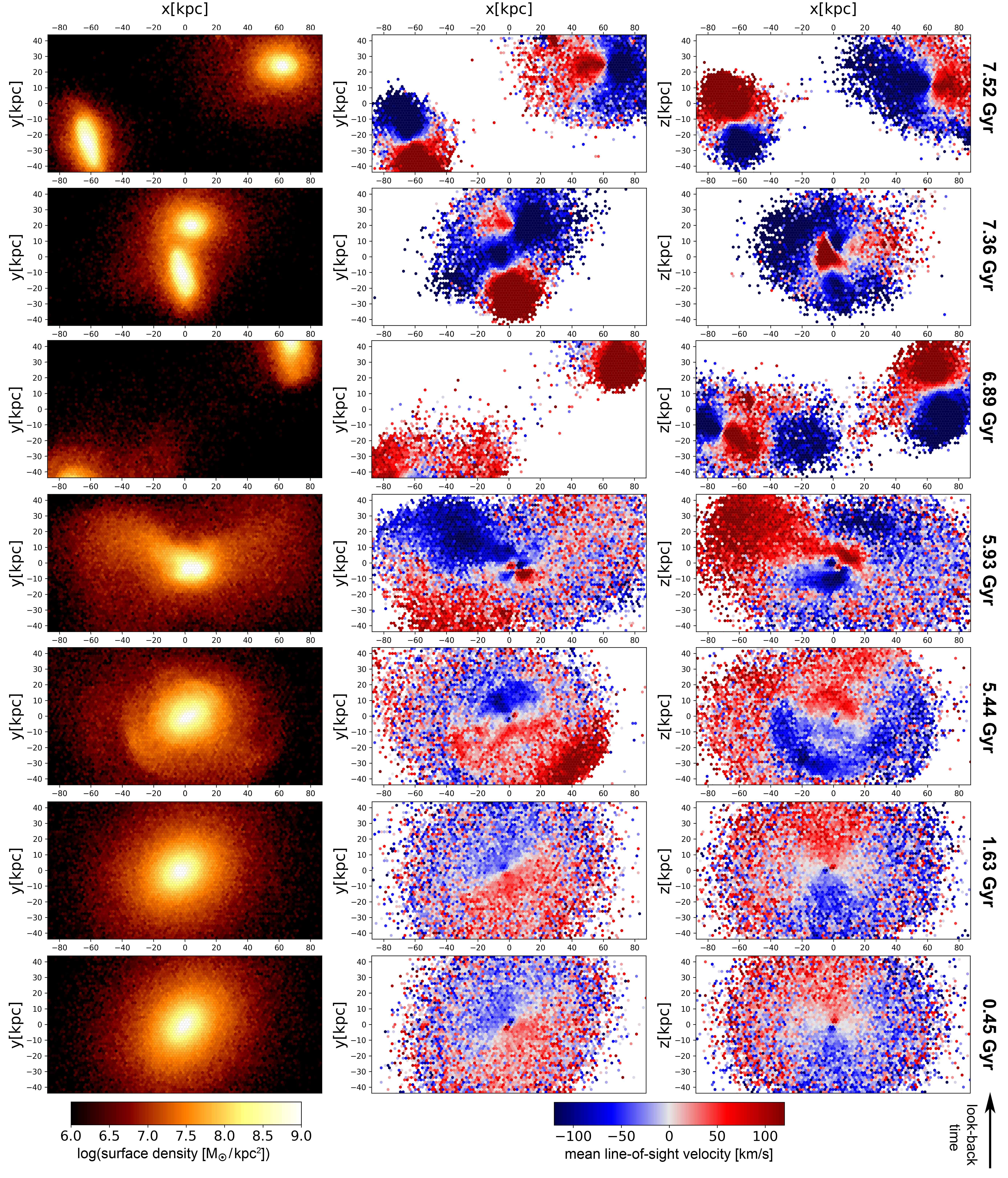}
\caption{
Snapshots of the major merger and the subsequent evolution of the galaxy \#140593. 
The $xy$ corresponds to the collision plane. 
Left column: the surface density; 
middle and right column: the mean line-of-sight velocity in two different projections. 
The look-back time for each snapshot is indicated on the right.
The size of each box is $176\times88$\,kpc.
\label{fig:dekin}
}
\end{figure*}

The galaxy \#140593 (the top left panels of Figs.~\ref{fig:ex}, \ref{fig:cos12}, and \ref{fig:shist}) underwent ten mergers 12.2\,--\,9.2\,Gyr ago.
After three quiescent gigayers, it suffered a major, 1:1, merger 6.1\,Gyr ago and the KDC appeared just one snapshot later.
Fig.\,\ref{fig:dekin} captures different stages of the merger. 
Exceptionally, in this figure, the $xy$ corresponds to the collision plane (not the principal axes of the galaxy).
The merger progenitors had the first pericentric passage 7.4\,Gyr ago and merged shortly after the second pericentric passage around 6\,Gyr.
In addition to the KDC, prolate rotation emerged during the merger as well. 
This galaxy was also a part of our sample of 59 Illustris prolate rotators published in \cite{illprol}.
The direction of the overall prolate rotation remains stable in the simulation box for the whole 6\,Gyr after the merger.
On the other hand, the direction of the rotation axis of the KDC slowly changes causing the KDC to disappear for a while in some projection planes. 
For example, the KDC is visible in the merger plane (the middle column of Fig.\,\ref{fig:dekin}) after the merger and also at the end of the simulation, but disappears around the look-back time 1.63\,Gyr.
However, any time after the merger, there are planes in which the KDC is well visible and distinct for the whole 6\,Gyr.

\begin{figure*} [!htb]
\centering
\includegraphics[width=\hsize]{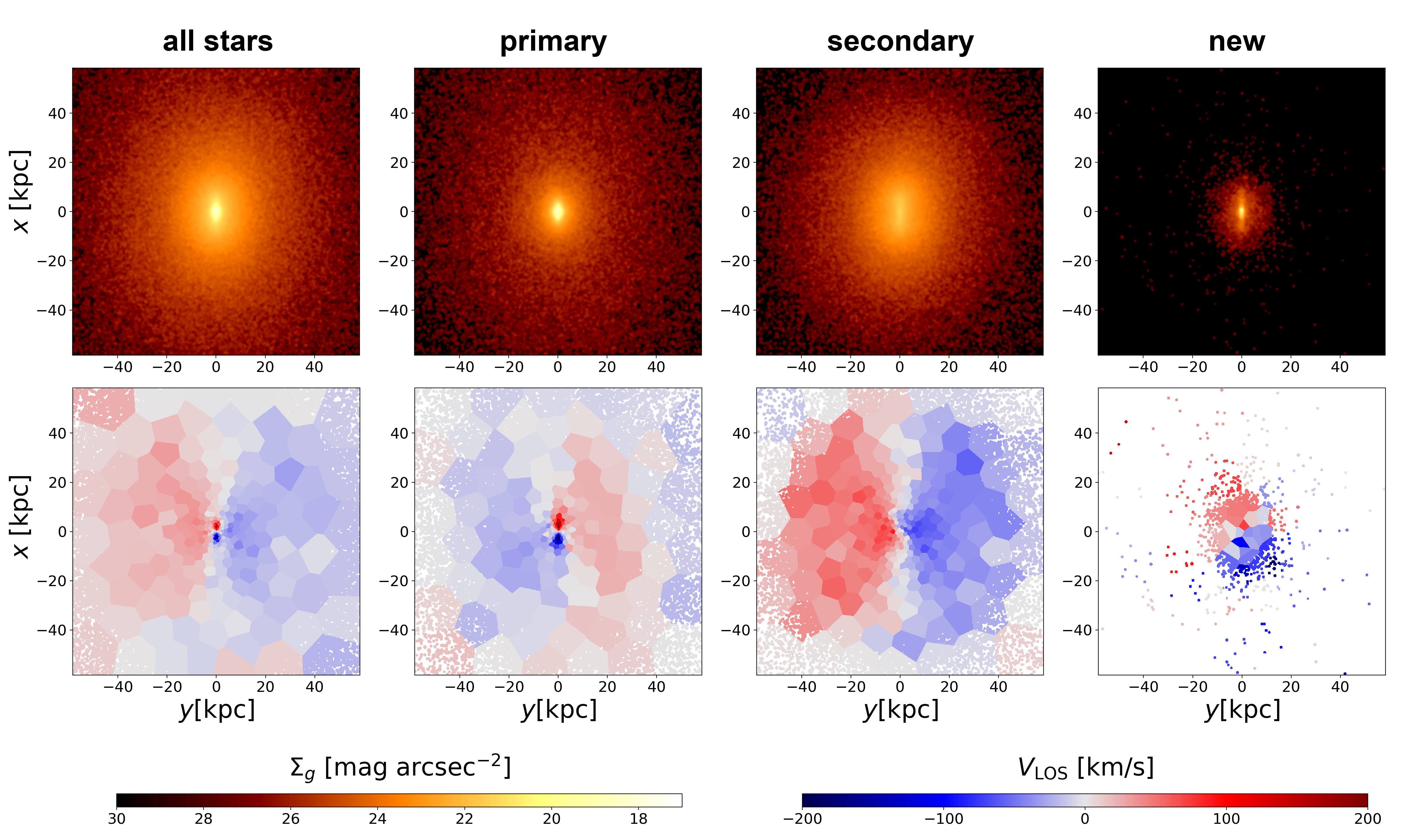}
\caption{
Maps of the stellar surface brightness and the mean line-of-sight velocity for the galaxy \#140593 viewed along the 3D minor axis in the last snapshot, $z=0$, divided into components according to their origin. 
The fields of view correspond to $1.5\times1.5\,r_{\rm max}$.
\label{fig:apsn}
}
\end{figure*}

Fig.\,\ref{fig:apsn} shows the situation in the galaxy \#140593 at the end of the simulation, $z=0$, in the same projection plane, divided into components according to their origin. 
Maps in the first column include all stellar particles;  
the second (third) column includes only particles that were already present in the
primary (secondary) progenitor before the merger; 
and the last column takes into account only stellar particles formed after the merger. 
We see that the large-scale prolate rotation is ensured by the particles originated in the secondary progenitor, despite the fact that the particles from the primary display a large-scale prolate rotation in the opposite sense. 
The KDC consists mostly of the old particles from the primary that managed to maintain the original disky rotation. 
The end of the merger was accompanied by the black hole activity and a significant gas loss (about three quarters of the total gas amount).
Star formation was considerably suppressed during and after the merger (see also the top left panel of Fig.\,\ref{fig:shist}), but some stars (1.7\,\% of the total stellar matter) formed after the merger to mostly support the KDC rotation. 

The galaxy \#353280 (the bottom left panels of Figs.~\ref{fig:ex}, \ref{fig:cos12}, and \ref{fig:shist}) underwent a relatively small, 1:6, merger 4.5\,Gyr ago and the KDC appeared shortly after that. 
The KDC rotation is partially supported by the stars formed in the primary progenitor prior to the merger, but it mostly consists of stars formed during the period of increased star formation after the merger. 
At the time of the KDC birth, the primary was a gas-rich galaxy with the gas fraction of 0.6 before and 0.5 after the merger. 
However, there was a 2-Gyr period of lower star formation prior to the KDC formation, possibly due to the ongoing merger, and it was not until the end of the merger when a significant amount of stars formed with angular momenta supporting the KDC rotation.
The KDC even survives a subsequent, more major, 1:3, merger 0.5\,Gyr before the end of the simulation. 

\begin{figure*} [!htb]
\centering
\includegraphics[width=\hsize]{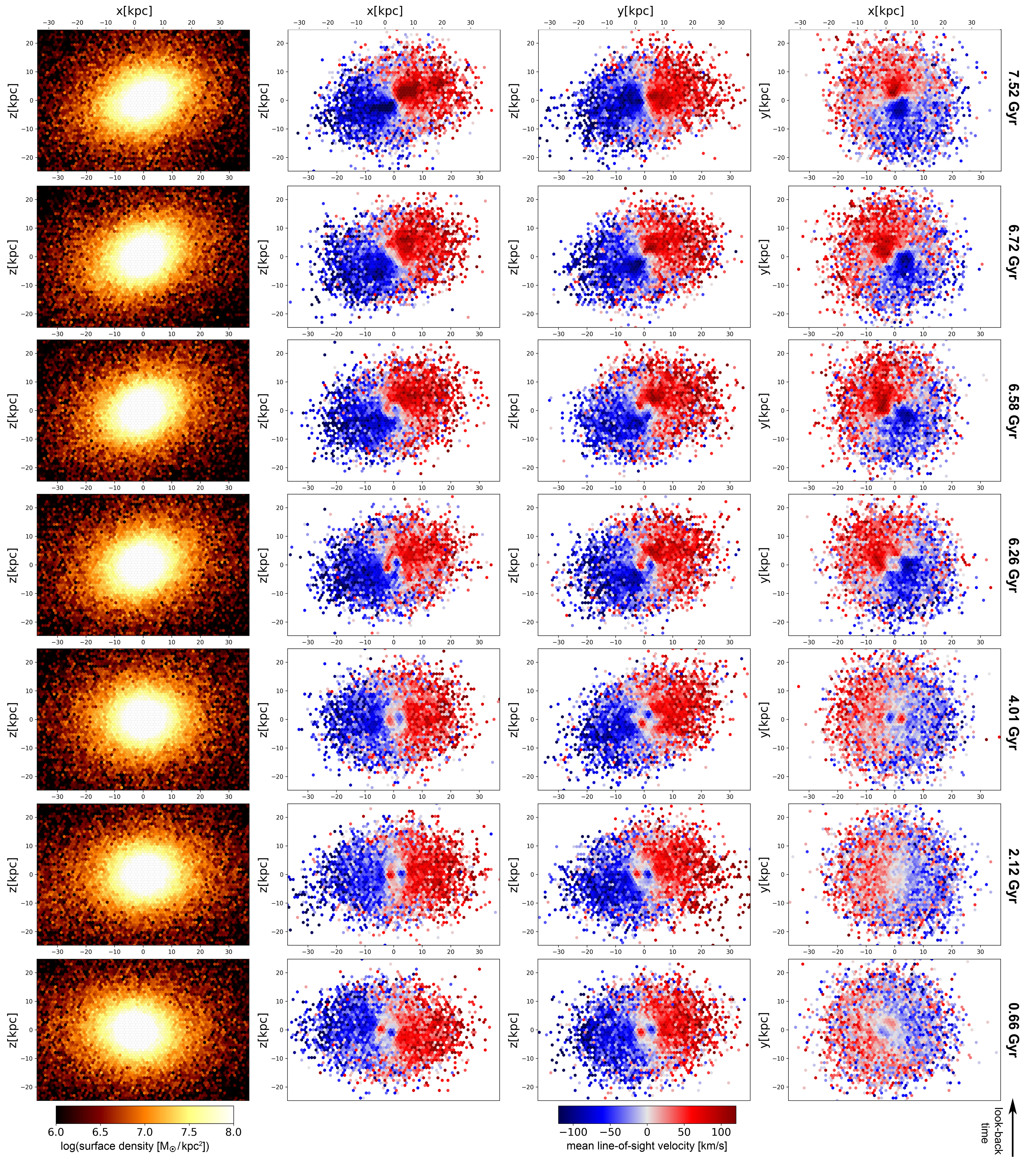}
\caption{
Snapshots of the evolution of the galaxy \#354838, $x$, $y$, and $z$ corresponds to the major, intermediate, and minor axis of the galaxy at the end of the simulation. 
The look-back time for each snapshot is indicated on the right.
The fields of view correspond to $1.5\times1.0r_{\rm max}$.
\label{fig:snaps}
}
\end{figure*}

The galaxies \#394799 and \#354838 (the top and bottom right panels, respectively, in Figs.~\ref{fig:ex}, \ref{fig:cos12}, and \ref{fig:shist}) experienced a series of mergers, but, in both cases, the last merger happened more than 3\,Gyr before the KDC birth.
At the time of the KDC birth, the galaxy \#394799 is very isolated -- no subhalo more massive than 10\,\% of the galaxy (in baryonic component) to the distance at least 1.6\,Mpc. 
The KDC rotation is supported mostly by stellar particles formed about 3.5\,Gyr before the KDC was born.
The galaxy \#354838 experienced the first pericentric passage of a smaller subhalo (about 20\,\% of the baryonic mass of the primary)in the distance of about 130\,kpc at the time of the KDC birth. 
The subhalo continues to orbit the galaxy and is about to merge at the end of the simulation, more than 6\,Gyr after the first pericenter passage.
After the KDC was born, the black hole in the host galaxy went through an active phase accompanied by a significant gas loss (change of the gas fraction from 0.5 to 0.3)  and the star formation was suppressed to a very small amount (0.55\,\% of the host stellar mass is younger than 5\,Gyr) within about 1\,Gyr after the KDC birth. 
Nevertheless, a large portion of stars formed during this 1-Gyr period supports the KDC rotation together with a large number of stars 1--3\,Gyr older than the KDC.
Fig.\,\ref{fig:snaps} shows a selection of snapshots from the evolution of the galaxy galaxy \#354838. 
The KDC emerged quite abruptly and it was visible in all projection planes in look-back times 6.4\,--\,2.4\,Gyr. 
After that, it disappeared in the $xy$ plane but emerged again after about 0.5\,Gyr.
The orientation of the projection planes is derived in the last snapshot, but kept the same for all outputs, as described in Sect.\,\ref{sec:birth}.

\section{Discussion} \label{sec:dis}

We examined our sample of 134 galaxies with KDCs visually selected from the last snapshot of the Illustris simulation.
Due to the limited Illustris resolution of individual galaxies, our sample covers larger diameters of kinematically distinct components than most of the KDCs traditionally identified in observations. 
\cite{mcd06} reported KDC radii 0.1\,--\,1.6\,kpc in the SAURON and OASIS data. 
\cite{kz00} found typical KDC radii of 0.4\,kpc, while KDC in \cite{meh98} and \cite{hau06} have sizes 1.4\,--\,6.4\,kpc.
The smallest KDC radius in our sample is 2.5\,kpc, half of our KDCs has $r_{\rm KDC}<10$\,kpc, but two of the biggest KDCs reach up to 35\,kpc radius.
However, the properties of the KDCs and their hosts as well as their formation histories do not seem to depend on the relative sizes of the KDCs (Sects.~\ref{sec:age} and \ref{sec:corr}, Figs.~\ref{fig:sa}, \ref{fig:agebir}, and \ref{fig:gas}).
Thus, including more extended KDCs does not skew our statistics on the Illustris KDCs.

Generally, the Illustris KDCs and their host galaxies are a very diverse group. 
For the well-resolved part of the sample (above host stellar mass $7.4\times10^{10}$\,M$_{\sun}$), the host masses follow the general distribution of the Illustris galaxies, with a possible slight preference towards more massive galaxies (Sect.\,\ref{sec:selection}, Fig.\,\ref{fig:Mdist}). 
This is in contrast with another kinematical peculiarity -- prolate rotation, which shows a clear preference for more massive galaxies in the observations \citep{tsa17,a3d2} as well as in the Illustris simulation \citep{illprol}.
In the Illustris simulation, the vast majority of prolate rotators originate in mergers. 
Also merger signs are more frequent among the high-mass galaxies \citep{tal09,matlas20}.
Thus, the mass distribution of the KDC hosts alone indicates that the KDCs are an inherently different group of kinematical peculiarity and that their formation mechanism must be, at least in a significant part, different from the prolate rotation. 
Indeed, only about half of the KDC origins can be connected directly to a merger event (see below). 
Moreover, compared to the prolate rotation, a lot of KDCs emerged quite early (see below). 
Early on, the mergers were much more frequent for all types of galaxies and the other signs of these early mergers (apart from the KDCs) are mostly undetectable at the redshift $z=0$.

In the well-resolved part, we detect KDCs in 8\,\% (90 out of 1160) of Illustris galaxies. 
We count all KDCs that are visible in at least one of the projections on the three principal planes of the host galaxy. 
For the whole sample of 134 galaxies, 44, 45, and 45 KDCs are visible in one, two, and three planes, respectively, making the KDCs, on average, visible almost exactly in two projections out of the three possibilities. 
That implies a roughly $8\times\frac{2}{3}\approx5$\,\% detection rate in the observation among all galaxies above the stellar mass $7.4\times10^{10}$\,M$_{\sun}$. 
The observed KDC incidence among early-type galaxies is around 13\,\% in the ATLAS$^{3{\rm D}}$ complete volume- and magnitude-limited sample of 260 ETGs and it ranges from 2 to 33\,\% in other observed samples of ETGs (see Sect.\,\ref{sec:intro}). 
The morphological types of the Illustris galaxies are not unequivocally determined, but if we assume that all KDC hosts in our sample are ETGs and that ETGs account for 30\,\% of all galaxies, the Illustris KDC sample would imply a 17\,\% detection rate among the ETGs. 
However, since our range of KDC sizes differs from the range in observed samples, this percentage is not easily compared with the available observations. 

KDCs can be long-lived and resilient features. 
The times of their births are distributed roughly uniformly between look-back times 0 and 11.4\,Gyr (Sect.\,\ref{sec:birth}, Fig.\,\ref{fig:merbir}). 
From our sample, 35 KDCs (26\,\%) survived a subsequent merger or multiple mergers (with stellar-mass ratio at least 1:15) that happened at least 1\,Gyr after the KDC birth. 
These subsequent mergers or flybys work both ways -- some of them make the KDC more prominent or compact, some make it more extended or weaker. 

There is no single channel of KDC formation, but mergers seem to be the main formation mechanism -- 81 (60\,\%) KDCs have an associated merger (Sect.\,\ref{sec:merger}). 
The association between mergers and KDC birth is especially strong for more massive hosts (the top left panel of Fig.\,\ref{fig:agebir}) and for major mergers (Fig.\,\ref{fig:mrat}). 
In the major-merger simulations of \cite{hof10}, KDCs are found only in remnants with the initial gas fraction at least 0.15 and most of them in 0.15\,--\,0.2.
In our sample, there are only two KDCs with initial gas fraction around 0.05, other associated mergers have at least 0.2, and there are also plenty of very gas-rich associated mergers (the left panel of Fig.\,\ref{fig:gas}).
The gas fraction at the time of the KDC birth (the right panel of Fig.\,\ref{fig:gas}) is, on average, about 20\,\% lower than the initial merger gas fraction. 

The KDCs without associated mergers can be formed during earlier pericentric passages of the merger progenitors, but the passages happened too soon before the merger for it to count as an associated merger. 
Others formed during a flyby of another galaxy, which did not lead to a merger event.
Some hosts formed (usually during mergers) rapidly rotating cores, which have angular momenta aligned with the global host rotation, and only later on, the core changes the orientation of the rotation axis. 
This can happen after another merger and in this case, the KDCs falls into the category \textit{associated mergers}.
Other KDCs form without an obvious external cause. 
There are roughly between 9 and 25 KDCs formed in a time period without a merger, pericentric passage, or flyby, depending on the limits on the galaxy masses (or mass ratios), distance, and allowed time difference between the event and the KDC birth. 
These KDCs are more frequent among lower-mass hosts as expected from the fact that, generally, lower-mass hosts are less likely to have the KDC birth associated with a merger (see the top left panel of Fig.\,\ref{fig:agebir}). 

\cite{dd16} proposed that some observed KDCs can be warped disks observed in face-on projections.
Generally, this phenomenon cannot account for a large portion of our sample, since, even though we specifically examined face-on projections, our KDCs are more likely to be found in the edge-on projections. 
KDCs are detected in the $zy$-plane twice as often as in the $xy$-plane, while the detection rate in $zy$ and $xz$-plane is about the same.
Nevertheless, we found one or two KDCs that can actually result from such a type of mechanism. 
These KDCs also show a change of their rotation axis, which can reflect the evolution of the warp. 

The change of the KDC rotation axis is a common phenomenon among the Illustris KDCs. 
Some degree of more or less regular precession occurs in more than a half the KDCs for at least a part of their lifetime, regardless of the KDC origin (with or without the associated merger). 
\cite{sch17} reported the precession for their KDC formed in a simulation of equal-mass spiral progenitors. 
This KDC dissolved after about 3\,Gyr reportedly due to the precession motion. 
In our Illustris KDC sample, we do not see any correlation between the precession and the KDC lifespan. 
There are six KDCs with especially pronounced precession in our sample. 
Four of them emerged during mergers, with close-to-equal mass ratios, between 5.3 and 10\,Gyr ago.
The precession is evaluated only in the periods without significant mergers, in which the global rotation and orientation of the galaxy is stable in the simulation box or changing very slowly compared to the KDC precession.

There is basically no correlation between the time of the merger closest to the KDC birth  and the mean mass-weighted stellar age of the KDC (the bottom right panel of Fig.\,\ref{fig:agebir}), but the relation between the KDC birth and KDC stellar age exists (the bottom left panel of Fig.\,\ref{fig:agebir}).
In both groups -- with or without the associated mergers (Sect.\,\ref{sec:corr}) --  we found  the stellar age to be roughly the same as the look-back time of the KDC birth for a quarter of the galaxies; the rest of the galaxies show an older stellar age than the look-back time of the KDC birth. 
This means that even though KDCs can be long lived features, an old stellar population inside the KDC radius does not automatically mean that the KDC exists there for a long time. 
On the other hand, younger KDC stellar ages indicate recently-born KDCs.

The existence of the relation between the KDC birth and KDC stellar age can be partially explained by the fact that the KDC birth is also often associated with a sudden loss of a significant amount of gas in the host galaxy.
For KDCs with the stellar age within $\pm1$\,Gyr from the KDC birth, 58\,\% (20 out of 34) have an associated sudden gas loss.
The remaining 100 KDCs have stellar age more than 1\,Gyr older than the birth and 30\,\% of them have a gas loss event associated with the KDC birth.
From these 50 cases (37\,\%) of associated gas loss, 37 happened during mergers (i.e. 46\,\% of the KDCs with associated mergers).
For the KDCs without associated mergers, 13 KDCs (25\,\%) have the KDC birth associated with the gas loss event. 
These events can be connected to the AGN activity or flybys of other galaxies.

\begin{figure*} [!htb]
\begin{center}
\includegraphics[width=0.75\hsize]{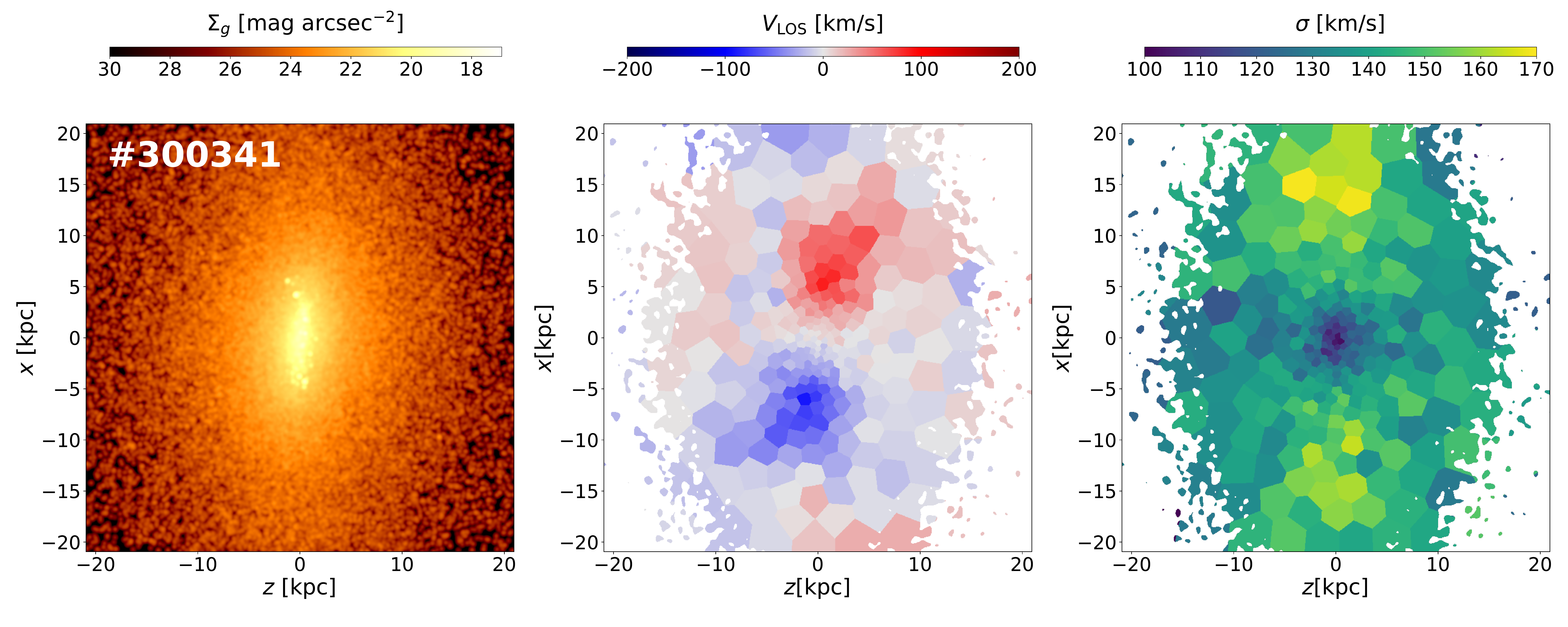}\\
\includegraphics[width=0.75\hsize]{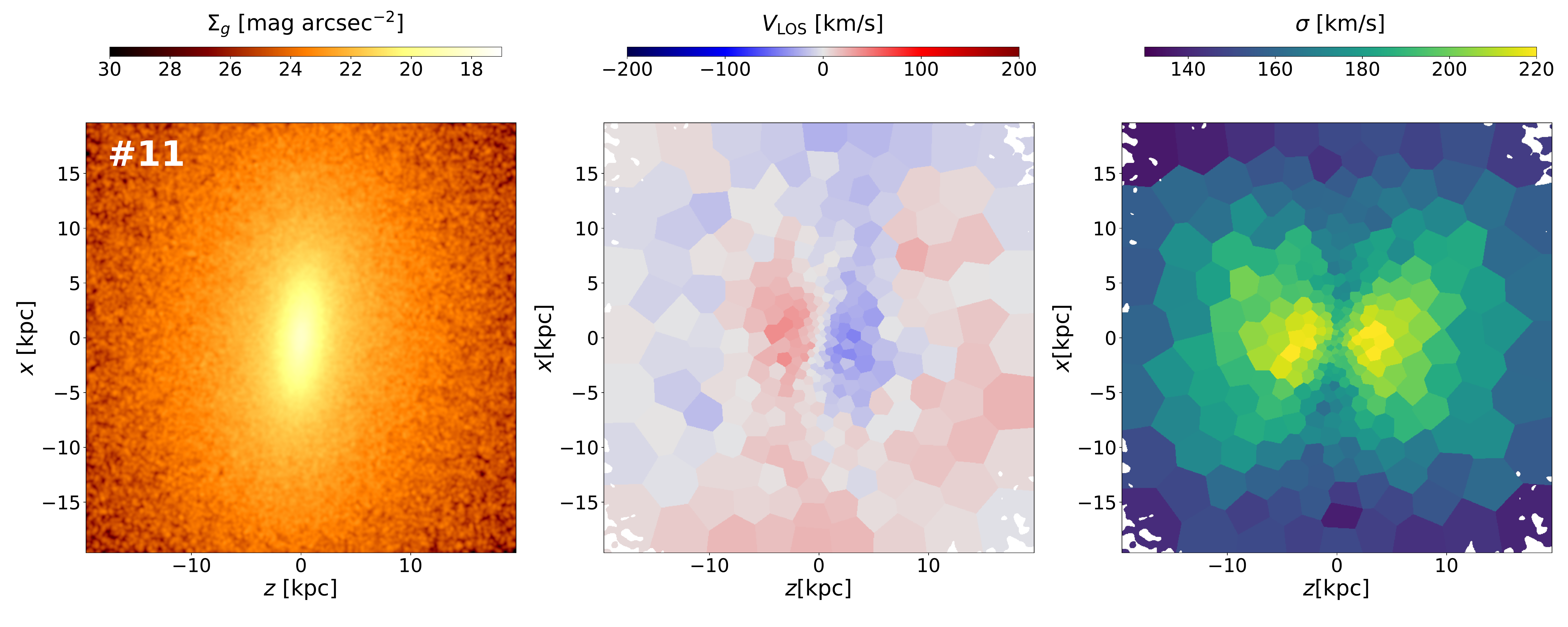}
\end{center}
\caption{
Examples of two Illustris KDC hosts that show two peaks in their maps of the velocity dispersion. 
The fields of view correspond to $6\times6\,r_{\rm eff}$.
\label{fig:dispex}}
\end{figure*}

Even though our KDC sample covers a different range of KDC sizes, it partially recovers the size\,--\,age trend reported by \cite{mcd06} for the KDCs in the SAURON sample -- KDCs with older stellar ages cover a wide range of sizes, while younger stars are associated with smaller $r_{\rm KDC}$ (the first panel of Fig.\,\ref{fig:sa}).
However, in our case, the trend mostly disappears when the graph is corrected for the relative sizes of the KDCs (the middle panel of Fig.\,\ref{fig:sa}).

In the ATLAS$^{3{\rm D}}$ sample, around 30\,\% of KDC hosts are classified as \textit{Double $\sigma$} ($2\sigma$) galaxies. 
This feature is thought to be a signature of two counter-rotating, co-spatial components \citep{a3d2} and have been reproduced in merger simulations \citep{jess07,a3d6}. 
\cite{a3d2} defined $2\sigma$ as two off-center, but symmetric, peaks in the velocity dispersion, which lie on the major axis of the galaxy, with the distance between the peaks at least half the effective radius. 
They found them in the ATLAS$^{3{\rm D}}$ sample by visually inspecting velocity dispersion maps.
We performed the visual inspection of the velocity dispersion maps of our sample of Illustris KDC hosts and found between 23 (counting only clear cases) and 50 (including also likely and possible cases) $2\sigma$~peak galaxies, i.e. 17\,--\,37\,\%. 
In the vast majority of cases, the $2\sigma$ peaks are recorded in the same projection plane (or planes) as the KDCs. 
Only 43\,\% of the clear cases of $2\sigma$ KDC hosts have associated mergers.
The top row of Fig.\,\ref{fig:dispex} shows the galaxy \#300341 -- an example of a $2\sigma$~galaxy among our Illustris KDC hosts. 
We found additional 12\,--\,15 galaxies that have symmetric peaks in the velocity dispersion lying on the minor axis. 
Nine of these galaxies show prolate or prolate-like large-scale rotation, out of which six have associated mergers. 
The bottom row of Fig.\,\ref{fig:dispex} shows the galaxy \#11 -- an example of an Illustris KDC host with large-scale prolate rotation and two peaks of the velocity dispersion on the minor axis.

\begin{figure*} [!htb]
\resizebox{\hsize}{!}{\includegraphics{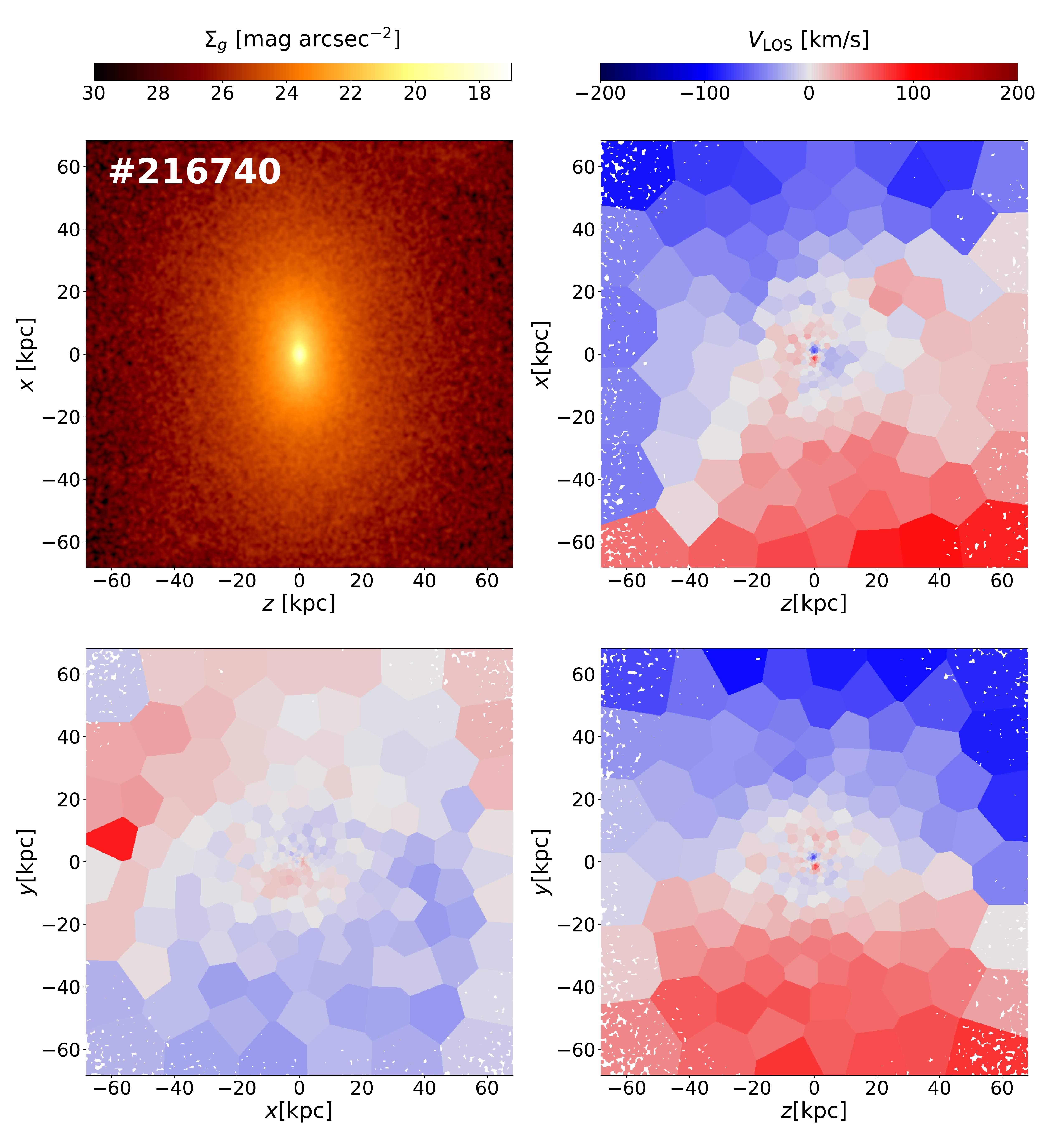}\hspace{3 cm}\includegraphics{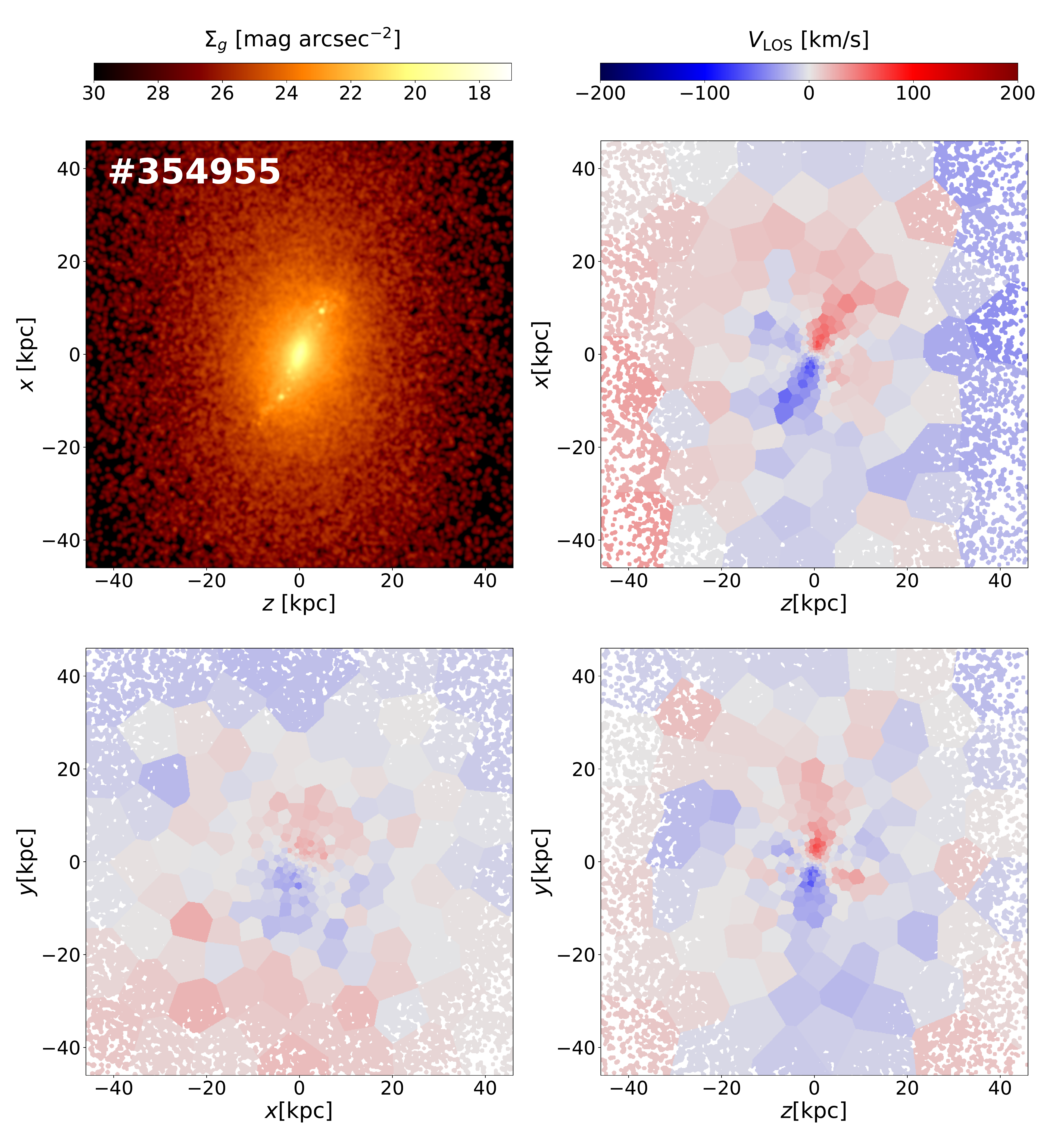}}
\caption{
Examples of two Illustris galaxies with a complicated kinematic structure of the KDCs. 
The fields of view correspond to $1.5\times1.5\,r_{\rm max}$. 
\label{fig:doubex}}
\end{figure*}

Around 20 KDC hosts show some kind of ring or oval shaped structure in the velocity dispersion maps. 
Interestingly enough, there are four cases of counter-rotating disks in our KDC sample and all four galaxies show this ring/oval feature. 
There are two KDC that show a complicated structure in the mean line-of-sight velocity fields.
The galaxy \#216740 -- the left part of Fig.\,\ref{fig:doubex} -- seems like a KDC nested in a larger KDC.
This qualitatively resembles the inner structure of NGC\,4528 \citep[see Figure\,C5 in][]{a3d2}.
The \#216740 galaxy underwent two merging periods. 
The first one was 7.7\,--\,6.9\,Gyr ago with two mergers, 1:1.5 and 1:5.0, after which the galaxy developed prolate rotation with a rapidly rotating core. 
The core maintained the rotation aligned with the host for about 4\,Gyr and became decoupled 3.2\,Gyr ago during a penultimate pericenter passage of a galaxy, with the mass ratio 1:1.7, that merged with the host 2.7\,Gyr ago.
After the merger, the host regained more oblate-like rotation and the KDC acquired the nested appearance.
The other galaxy with a complicated KDC structure, \#354955 -- the right part of Fig.\,\ref{fig:doubex} -- appears to have two perpendicular disks in its center.
Its KDC emerged 8.6\,Gyr ago without an associated merger. 
After a series of mergers (1:11, 1:1.4, and 1:3.1) 6.2\,--\,4.5\,Gyr ago, the KDC gained the disky shape and gradually changed the rotation-axis direction by at least $90\degrees$ between the last merger and the end of the simulation. 

The presence of a KDC may be correlated with other morphological features indicating a possible past merger in the history of the galaxy, such as is the case of the so-called shell galaxies. 
Shell galaxies are mostly early-type galaxies that possess, in their stellar component, a low surface brightness structure in the form of concentric arcs.
Shell galaxies have been believed to result from close-to-radial minor mergers for decades
\citep{q84, dc86, hq88, e12sg}, but recently the view concerning their origin has been shifting from minor mergers to intermediate-mass \citep{duc15,e20sg} or even major ones \citep{illsg17}.

In the nineties, there was a suspected association of KDCs and the stellar shells.
As stated in \cite{hau99}, the strong association comes from the finding by \cite{for92} that, at the time, all of the nine well-established KDCs and a further four out of the six `possible KDCs' hosts possess the stellar shells. 
In the current observational data, the association is much weaker -- from the 38 KDCs of the ATLAS$^{3{\rm D}}$ and SAURON sample \citep{mcd06,saur9,a3d2}, to our knowledge, only 12 hosts (31\,\%) are known shell galaxies. 
Moreover, \cite{mp04} examined a sample of 117 ellipticals and found KDCs at a comparable rate (around 16\,\%) in the groups of galaxies with and without morphological peculiarities. 

We found 71 (53\,\%) shell galaxies in our sample of Illustris KDCs, although some of the shell galaxies can be reliably identified only in the images with a field of view larger than $2\times2\,r_{\rm max}$.
For example the galaxy \#140593 has the shells well visible in the inner parts shortly after the merger (see the snapshot at look-back time 5.44\,Gyr in Fig.\,\ref{fig:dekin}), but at the end of the simulation, the shells are still clearly visible in the outer parts of the galaxy.
Naturally, not all these shells are connected to the KDC origin, since there are 11 (21\,\%) shell galaxies in the group of the KDC hosts without associated mergers.
In the group with associated mergers, 60 (74\,\%) KDC hosts possess shells. 
Incidence of the shells is the highest for the KDCs recently created in mergers -- of all 29 KDC hosts that had an associated merger during the last 4.5\,Gyr, all but one have visible stellar shells.

We compared these findings with the sample of 59 Illustris prolate rotators \citep{illprol} -- since the vast majority of them was created in a merger during the last 6\,Gyr of the simulation, we expect a higher rate of the shell galaxies. 
We inspected the surface-brightness maps of the prolate rotators (constructed in the same way as for the KDC hosts) and found 40 (68\,\%) shell galaxies. 
This is not much higher than the KDC sample, but the prolate rotators have a higher share of low-mass galaxies, since the prolate rotation is easier to detect than KDCs in galaxies with lower numbers of particles. 
However, the shells are not easily recognized in the low-resolution data and dissolve faster in low-mass galaxies.
In the well-resolved mass (above $7.4\times10^{10}$\,M$_{\sun}$) shells are found in 74\,\% (67 out of 90) of KDC hosts and 94\,\% (17 out of 18) of prolate rotators.
Based on the Illustris data, we predict the incidence of the shells (or tidal features in general) to be higher in prolate rotators than in the KDC hosts.

\section{Conclusions} \label{sec:con}

We used the publicly available data of the large-scale cosmological simulation Illustris that follows a coevolution of dark and baryonic matter.
We visually inspected kinematical maps of all 7697 galaxies that have more than $10^4$ stellar particles in the final output (redshift $z=0$) of the Illustris-1 run. 
We identified 134 galaxies with kinematically distinct cores (KDCs).

We determined look-back times of the KDC births and analyzed merger histories of the KDC hosts. 
We divided our sample into two groups according to the KDC origin: 
(1) 81 galaxies that have the KDC birth associated with a merger event, and 
(2) 53 galaxies without an associated merger.

We summarize our findings as follows: 
\begin{itemize}
\setlength\itemsep{0.3em}
\renewcommand{\labelitemi}{$\bullet$}
	\item equal portions of the KDCs are detectable in one, two, and three projection planes with the least probability of detection in the face-on projection; 
\item for the well-resolved galaxies, the host masses follow the general distribution of the Illustris galaxies, with a possible slight preference towards more massive galaxies; 
\item KDCs can be long-lived features -- their births are distributed roughly uniformly between look-back times 0 and 11.4\,Gyr; 
\item there is no single channel of KDC formation; 
\item mergers seem to be the main formation mechanism -- 60\,\% KDCs emerged during (or shortly after) a merger event; 
\item other KDCs formed during a pericentric passage or flyby of another galaxy, by a precession of previously formed rapidly rotating core, or without an obvious external cause; 
\item the association between mergers and KDC birth is especially strong for more massive hosts and for major mergers; 
\item the mean mass-weighted stellar age of the KDC is either about the same as the look-back time of the KDC birth or older, i.e. an old stellar population of the KDC does not automatically imply an early KDC formation; 
\item one quarter of KDCs have similar values of the KDC stellar age and the KDC birth time (regardless of the KDC origin); 
\item we partially recover the KDC size\,--\,age trend reported by \cite{mcd06}, but it mostly disappears when corrected for the KDC relative sizes;  
\item the KDC birth is associated with a sudden loss of gas of the host for 46\,\% of the KDCs with associated mergers and 25\,\% KDCs without ones; 
	\item KDCs can survive even major or multiple subsequent mergers; 
\item more than a half of the KDCs, regardless of their origin, experienced a precession of their kinematic axis, but the most pronounced precession occurs for KDCs that emerged during major mergers; 
\item around a quarter of the KDC hosts (with a mild preference towards KDCs without associated mergers) have the $2\sigma$ peaks in their velocity dispersion maps; 
\item there is a strong association between a KDC and stellar shells for the KDCs born in the associated merger during the last 4.5\,Gyr;
\item based on the Illustris data, we predict the incidence of the shells (or tidal features in general) to be higher in prolate rotators than in the KDC hosts. 
\end{itemize}

\begin{acknowledgements}
We thank the referee for valuable comments that helped improve the manuscript. This work was supported in part by the Polish National Science Centre under grant 2017/26/D/ST9/00449 (IE).
\end{acknowledgements}


\bibliographystyle{aa}
\bibliography{kdc} 


\end{document}